\documentclass[longauth]{aa}
\usepackage{graphicx}
\usepackage{natbib}
\usepackage{multirow}
\usepackage{hyperref}
\titlerunning{}

\usepackage{color}

\usepackage{cleveref}

\begin{document}

\title{VEGA/CHARA interferometric observations of Cepheids \\
I. A resolved structure around the prototype classical Cepheid $\delta$~Cep in the
  visible spectral range}
\titlerunning{VEGA/CHARA interferometric observations of Cepheids}
\authorrunning{Nardetto et al. }
\author{N. Nardetto \inst{1} 
\and A. M\'erand \inst{2} 
\and D. Mourard \inst{1}  
\and J. Storm \inst{3}  
\and W. Gieren\inst{4, 5}   
\and P. Fouqu\'e\inst{6}  
\and A. Gallenne\inst{2,4}  
\and \\ D. Graczyk \inst{4, 5, 7} 
\and P. Kervella \inst{8,9}  
\and H. Neilson\inst{10} 
\and G. Pietrzynski \inst{7}  
\and B. Pilecki \inst{7}  
\and J. Breitfelder \inst{8} 
\and P. Berio \inst{1}  
\and \\ M. Challouf \inst{1,11} 
\and J.-M. Clausse \inst{1}
\and R. Ligi \inst{1} 
\and P. Mathias\inst{12,13}
\and A. Meilland \inst{1} 
\and K. Perraut \inst{14,15} 
\and E. Poretti \inst{16} 
\and \\ M. Rainer \inst{16} 
\and A. Spang \inst{1} 
\and P. Stee \inst{1} 
\and I. Tallon-Bosc \inst{17} 
\and T. ten~Brummelaar\inst{18,19} 
}
\institute{Laboratoire Lagrange, UMR7293, Universit\'e de Nice Sophia-Antipolis, CNRS, Observatoire de la C\^ote d'Azur, Nice, France Nicolas.Nardetto@oca.eu   
\and European Southern Observatory, Alonso de C\'ordova 3107, Casilla 19001, Santiago 19, Chile 
\and Leibniz Institute for Astrophysics, An der Sternwarte 16, 14482, Potsdam, Germany 
\and Departamento de Astronom\'ia, Universidad de Concepci\'on, Casilla 160-C, Concepci\'on, Chile 
\and Millenium Institute of Astrophysics, Santiago, Chile
\and Observatoire Midi-Pyr\'en\'ees, Laboratoire d'Astrophysique, UMR 5572, Universit\'e Paul Sabatier - Toulouse 3, 14 avenue Edouard Belin, 31400 Toulouse, France  
\and Nicolaus Copernicus Astronomical Center, Polish Academy of Sciences, ul. Bartycka 18, PL-00-716 Warszawa, Poland. 
\and LESIA (UMR 8109), Observatoire de Paris, PSL, CNRS, UPMC, Univ. Paris-Diderot, 5 place Jules Janssen, 92195 Meudon, France 
\and Unidad Mixta Internacional Franco-Chilena de Astronom\'ia, CNRS/INSU, France (UMI 3386) and Departamento de Astronom\'ia, Universidad de Chile, Camino El Observatorio 1515, Las Condes, Santiago, Chile 
\and Department of Astronomy \& Astrophysics, University of Toronto, 50 St. George Street, Toronto, ON, M5S 3H4 
 \and Laboratoire Dynamique Mol\'eculaire et Mat\'eriaux Photoniques, UR11ES03, Universit\'e de Tunis/ESSTT, Tunisie 
   \and   Universit\'e de Toulouse, UPS-OMP, Institut de recherche en Astrophysique et Plan\'etologie, Toulouse, France
\and CNRS, UMR5277,  Institut de recherche en Astrophysique et Plan\'etologie, 14 Avenue Edouard Belin, 31400 Toulouse, France 
 \and   Univ. Grenoble Alpes, IPAG, F-38000 Grenoble, France
 \and   CNRS, IPAG, F-38000 Grenoble, France
  \and  INAF -- Osservatorio Astronomico di Brera, Via E. Bianchi 46, 23807 Merate (LC), Italy 
 \and Universit\'e de Lyon, Universit\'e Lyon 1, Ecole Normale Sup\'erieure de Lyon, CNRS, Centre de Recherche Astrophysique de Lyon UMR5574, F-69230, Saint-Genis-Laval, France 
 \and Georgia State University, P.O. Box 3969, Atlanta GA 30302-3969, USA  \and CHARA Array, Mount Wilson Observatory, 91023 Mount Wilson CA, USA }
\date{Received ... ; accepted ...}
\abstract{The B-W method is used to determine the distance of Cepheids and consists in combining the angular size variations of the star, as derived from infrared surface-brightness relations or interferometry, with its linear size variation, as  deduced from visible spectroscopy using the projection factor. The underlying assumption is that the photospheres probed in the infrared and in the visible are located at the same layer in the star whatever the pulsation phase. While many Cepheids have been intensively observed by infrared beam combiners, only a few have been observed in the visible.} {This paper is part of a project to observe Cepheids in the visible with interferometry as a counterpart to infrared observations already in hand.} { Observations of $\delta$~Cep itself were secured with the VEGA/CHARA instrument over the full pulsation cycle of the star.} {These visible interferometric data are consistent in first approximation with a quasi-hydrostatic model of pulsation surrounded by a static circumstellar environment (CSE) with a size of $\theta_\mathrm{CSE}=8.9\pm3.0$ mas and a relative flux contribution of $f_\mathrm{CSE}=0.07\pm0.01$. A model of visible nebula (a background source filling the field of view of the interferometer) with the same relative flux contribution is also consistent with our data at small spatial frequencies. However, in both cases, we find discrepancies in the squared visibilities at high spatial frequencies (maximum 2$\sigma$) with two different regimes over the pulsation cycle of the star, $\phi=0.0-0.8$ and $\phi=0.8-1.0$. We provide several hypotheses to explain these discrepancies, but  more observations and theoretical investigations are necessary before  a  firm conclusion can be drawn.} {For the first time we have been able to detect in the visible domain a resolved structure around $\delta$~Cep. We have also shown that a simple model cannot explain the observations, and more work will be necessary in the future, both on observations and modelling.}

\maketitle\

\section{Introduction}\label{s_Introduction}

The Baade-Wesselink (BW) method for the distance determination of Cepheids is, in its first version, purely spectro-photometric \citep{lindermann18,baade26, wesselink46}. Results by \citet{fouque07} and \citet{storm11a, storm11b} illustrate the central and current role of this method in the distance scale calibration, even if recent calibrations of the Hubble constant rely exclusively on the trigonometric parallaxes of a few Galactic Cepheids \citep{riess11, benedict07}. The first interferometric version of the BW method was attempted in the visible by \citet{mourard97}, and soon afterward in the infrared by \citet{kervella99, kervella01} and \citet{lane00}. Since then, the infrared interferometric BW method has been applied to a significant number of Cepheids, twelve in total (see Table~\ref{Tab.hra}), while among these stars only four have been observed by visible interferometers, and the pulsation could actually be resolved  for only one of them, $\ell$~Car \citep{davis09}. %

The principle of the interferometric version of the BW method is simple. Interferometric measurements lead to angular diameter estimations over the whole pulsation period, while the stellar radius variations can be deduced from the integration of the pulsation velocity. The latter is linked to the observational velocity deduced from spectral line profiles by the projection factor $p$  \citep{nardetto04,merand05, nardetto07, nardetto09}. 

There are several underlying assumptions to the BW method. First, the limb-darkening of the star is assumed to be constant during the pulsation cycle. This has no impact on the interferometric analysis, at least in the infrared \citep{kervella04a}, while in the visible \citet{davis09} reported the need to take into account a limb-darkening variation from Sydney University Stellar Interferometer (SUSI) observations. On the theoretical side, \citet{nardetto06b} found via hydrodynamical simulation that considering a constant limb darkening in the visible leads to a systematic shift of about 0.02 in phase on the angular diameter curve, which basically means that there is no impact on the derived distance (because the amplitude of the angular diameter curve is unchanged). Second, the projection-factor, mostly dominated by the limb-darkening calculated in the visible domain, is also assumed to be constant, which seems  reasonable, at least in theory \citep{nardetto04}. Third, when applying the BW method,  visible spectroscopy (e.g. \citealt{nardetto06a}) is often combined with infrared interferometric data, or even -- in the recent photometric version of the BW method -- with various photometric bands \citep{breitfelder15}. In the distance determination, we implicitly assume that the angular and linear diameters correspond to the same physical layer in the star.
In this context, testing these hypotheses using visible interferometric observations seems to be of prime importance. 
This can be done first by simply verifying the internal consistency of the BW distances derived from visible and infrared interferometry, and then comparing them with available precise parallaxes \citep{benedict07, majaess12}.

This paper is the first in a series which   investigates Cepheids with the Visible spEctroGraph and polArimeter (VEGA) beam combiner \citep{mourard09} operating at the focus of the Center for High Angular Resolution Astronomy (CHARA) array \citep{ten05} located at the Mount Wilson Observatory (California, USA). The study  focuses on the prototype $\delta$~Cep star. The VEGA data are first reduced (Sect.~\ref{s_vega}) and then  analysed in terms of uniform disk angular diameters (Sect.~\ref{s_UD}). In Sect.~\ref{s_CSE}, we show that $\delta$~Cep is clearly surrounded by a resolved structure, while some evidence in our interferometric data point toward an additional physical effect. We explore two hypotheses: a circumstellar reverberation and a strong limb-darkening variation. We draw our conclusions in Sect.~\ref{s_C}.

\begin{table*}
\caption{\label{Tab.hra} Cepheids for which an interferometric BW method has been applied. For some studies corresponding to the first attempts (indicated by  asterisks) the pulsation could not be clearly resolved angularly leading to a mean value of the angular diameter of the Cepheid. In the visible domain, the BW method was applied successfully to only one Cepheid, $\ell$~Car, while in the infrared, among the eleven Cepheids the pulsation could be resolved for all stars, even X~Sgr, which is known to be atypical likely owing to shockwaves  travelling within the atmosphere \citep{mathias06}.}
\begin{center}
\setlength{\doublerulesep}{\arrayrulewidth}
\begin{tabular}{ll}
\hline
\hline
Cepheids                                                         & reference                                    \\
\hline
\multicolumn{2}{c}{visible} \\
$\delta$~Cep$^{\star}$      & \citet{mourard97}  \\
$\alpha$~UMi$^{\star}$, $\zeta$~Gem$^{\star}$, $\delta$~Cep$^{\star}$, $\eta$~Aql$^{\star}$   & \citet{nordgren00} \\
$\delta$~Cep$^{\star}$, $\eta$~Aql$^{\star}$  & \citet{armstrong01} \\
$\ell$~Car & \citet{davis09} \\
\hline
\multicolumn{2}{c}{H band} \\
$\zeta$~Gem & \citet{lane00} \\
$\zeta$~Gem, $\eta$~Aql & \citet{lane02} \\
$\kappa$~Pav & \citet{breitfelder15} \\
$\ell$~Car &   \citet{anderson16} \\
X~Sgr, W~Sgr, $\zeta$~Gem, $\beta$~Dor, $\ell$~Car &  \citet{breitfelder16} \\

\hline
\multicolumn{2}{c}{K band} \\
$\zeta$~Gem$^{\star}$ & \citet{kervella01} \\
X~Sgr$^{\star}$, $\eta$~Aql, W Sgr, $\zeta$~Gem$^{\star}$, $\beta$~Dor, Y Oph$^{\star}$, $\ell$~Car &\citet{kervella04a}  \\

 $\delta$~Cep & \citet{merand05} \\
Y Oph & \citet{merand07} \\
FF Aql, T Vul & \citet{gallenne12} \\
$\delta$~Cep, $\eta$~Aql & \citet{merand15} \\
\hline
\end{tabular}
\end{center}
\end{table*}

\begin{table*}
\caption{\label{Tab.cals}  List and properties of calibration stars selected with the {\it SearchCal}  software provided by the Jean-Marie Mariotti Center (JMMC)  \citep{bonneau06, bonneau11}.  T$_\mathrm{eff}$ is the effective temperature, $g$ the gravitation acceleration, R the magnitude of the calibrator in the Johnson R filter, $\theta_\mathrm{UD}$ the uniform disk angular diameter for the R filter of the Johnson photometric system. These parameters were adopted from \citet{lafrasse10} for all calibrators, except HD~195725 (C3) for which we used the SearchCal tool itself. C1 and C2 were used in this study  to calibrate the $\delta$~Cep VEGA/CHARA data (see Fig.~\ref{Fig.ft}), while C3 to C8 were used to test the robustness of the C1 and C2 calibrators (see Fig.~\ref{Fig.cals}).}
\begin{center}
\begin{tabular}{lcccccc}
\hline
\hline
Calibrator HD        & number &       Spectral type  &  T$_\mathrm{eff}$ & $\log$ g           &  R & $\theta_\mathrm{UD}$ (R band)  \\
                                 &                 &                                &                K              &  [cgs]     &       &       [mas]                                         \\
\hline
HD 214734            &   C1         &           A3IV            &  8600           &   4.2      &  5.073      &   0.327  $\pm$ 0.023                             \\
HD 213558            &   C2         &          A1V               &  9500         & 4.1        &   3.770    &  0.458 $\pm$   0.033                                 \\
\hline
HD 195725            &   C3         &       A7III  &  8000 & 3.3  & 4.050  &  0.617 $\pm$   0.044  \\
HD 182564            &   C4         &          A2IIIs             &     9380            & 3.4        &  4.550   &  0.377 $\pm$ 0.027                                \\
HD 211336            &   C5         &          F0IV                &  7300             &  4.3       &   3.920  &   0.714 $\pm$ 0.051                                 \\
HD 214454            &   C6         &          A8IV                  &  7500            &  4.3       &  4.410   &      0.581 $\pm$ 0.042                                      \\
HD 3360                &   C7        &            B2IV                 &   20900            & 3.9        &  3.740    &          0.284 $\pm$ 0.020                                             \\
HD 192907            &   C8         &          B9III                       &  10500           &  3.4       & 4.410   &                0.346 $\pm$ 0.025                                            \\
\hline
\end{tabular}
\end{center}
\end{table*}

\section{VEGA/CHARA observations of $\delta$~Cep}\label{s_vega}

The CHARA array consists of six telescopes of 1 meter in diameter, spread in a Y-shaped configuration, which offers 15 different baselines from 34 meters to 331 meters. These baselines can achieve a spatial resolution up to 0.3 mas in the visible. The interferometric observations were secured with the VEGA/CHARA instrument. The journal of observations is presented in Tables~\ref{Tab.log1}~to~\ref{Tab.log4}. Given the large angular size of $\delta$~Cep, we used only two short baselines, i.e. S1S2 and E1E2, with projected baselines ranging from about 27 to 31~m for S1S2 and from 52 to 66~m for E1E2.
The data were processed using the standard VEGA pipeline \citep{mourard09, mourard11, ligi13} considering different spectral bands from 3~nm (in high spectral resolution mode) to 20~nm (in medium spectral resolution mode), and with reference wavelengths ranging from 500~nm to 745~nm (the minimum and maximum wavelengths of these bands are given in Tabs.~\ref{Tab.log1}~to~\ref{Tab.log4}).  

In order to calibrate the squared visibilities, we considered the references stars HD~214734 (C1) and HD~213558 (C2). The first was used to calibrate the data of $\beta$~Cep presented in \citet{nardetto11a}. Several nights of VEGA observations in 2008 and 2009 were also devoted to controlling the quality of these two calibrators by comparing their transfer functions (defined as the ratio of the actual to expected squared visibility) with those of six other calibrators listed in Tab.~\ref{Tab.cals} (hereafter C3 to C8). The consistency among the eight calibrators is shown in Fig.~\ref{Fig.cals}, while the general quality of VEGA/CHARA data is illustrated in Fig.~\ref{Fig.ft}.

Cycle-to-cycle variations have never been detected for $\delta$~Cep, although they have been for a few other Cepheids \citep{anderson14, anderson16}. The data are thus recomposed into a unique cycle and the data corresponding to nights that are  close in pulsation phase are merged (see phases $0.180$, $0.526$, $0.627$ in Tabs.~\ref{Tab.log1}~to~\ref{Tab.log4}).  Delta Cephei was finally observed on 20 separate nights corresponding to 17 different pulsation phases. As the observations over a given night can be spread over several hours, we first derive the phase for each observation during the night and then calculate the average and the corresponding standard deviation. The standard deviations of the pulsation phases are rather low and range from 0.001 to 0.017, except for $\phi=0.999$ where we get  an error of $0.025$ because the star was observed at the very beginning and at the very end of the night. The pulsation phases can be found in Tab.~\ref{Tab.ang} together with their respective errors, the number of visibility measurements ($N$), and the baseline used.   
For each calibrated visibility, the statistical and systematic calibration errors are given separately. The systematic calibration errors, owing to the uncertainty on the estimation of the diameter of the reference star, were found to be negligible compared to the statistical values. Therefore, we only considered the statistical uncertainties in the model fitting. In a few cases, the statistical uncertainty was clearly underestimated. Thus, we fixed the uncertainty on the calibrated squared visibility to 0.05 \citep{mourard12a} for the following nights: 2012 September 23, 2013 October 26, 2013 November 26, 2014 July 02, 2014 July 05, and 2014 July 09.



\begin{figure}[htbp]
\begin{center}
\resizebox{1.0\hsize}{!}{\includegraphics[clip=true]{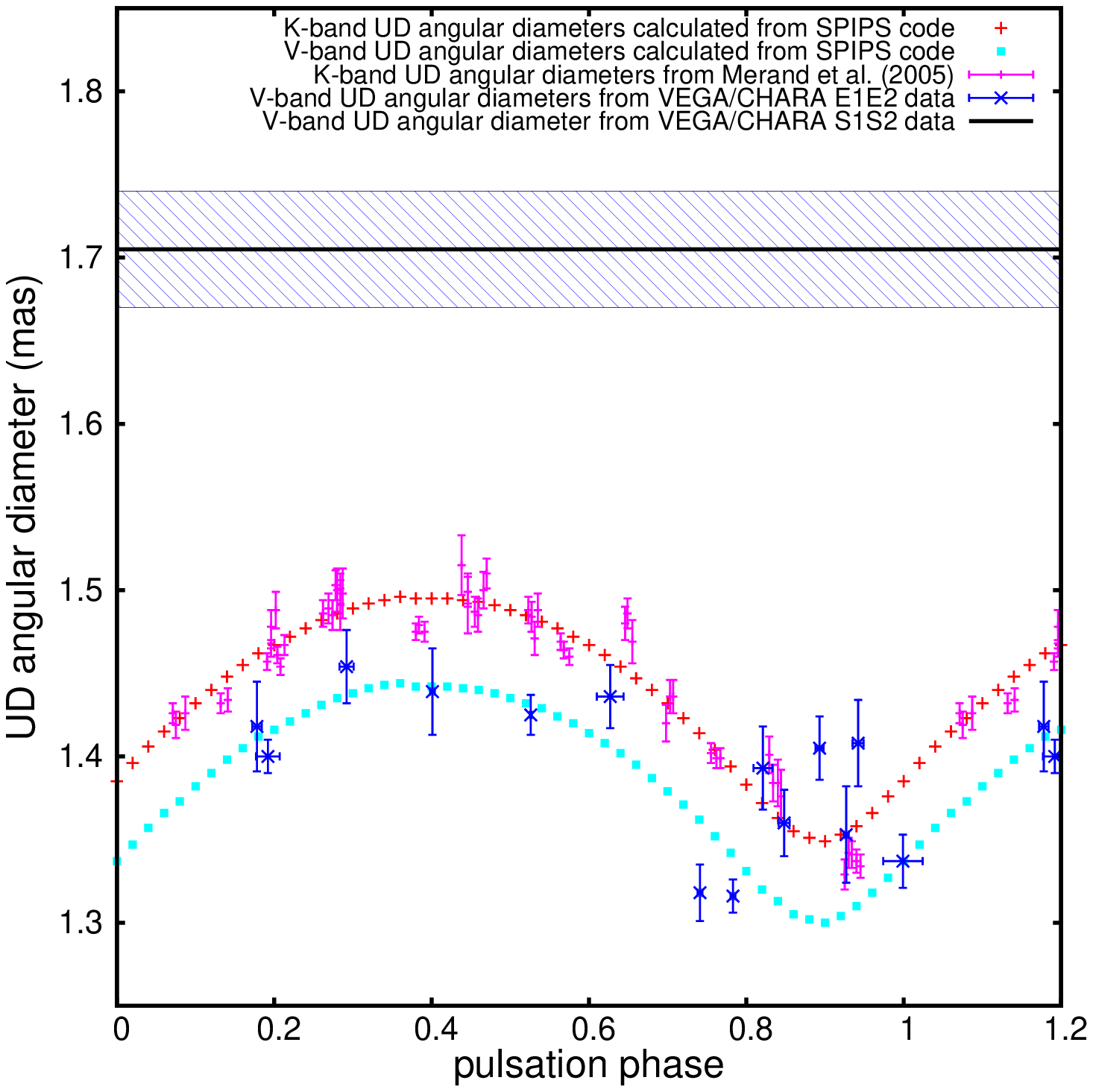}}
\end{center}
\caption{ Uniform disk diameters $\theta_\mathrm{UD}$ derived from the VEGA/CHARA E1E2 data are plotted as a function of the pulsation phase (blue crosses). The S1S2 data (corresponding to four different pulsation phases, see Tab.~\ref{Tab.ang}) were merged and the corresponding mean  $\theta_\mathrm{UD}$ is represented by a black horizontal line together with the $1\sigma$ uncertainty (blue dashed zone). The uniform disk angular diameter curve from FLUOR/CHARA in the K band \citep{merand05} is also shown for comparison. The red crosses and the light blue squares show the K band and R band UD angular diameter curves, respectively, as predicted from the SPIPS code \citep{merand15}.} \label{Fig.diam}
\end{figure}

\begin{table}
\caption{\label{Tab.ang} The pulsation phases, the number of visibility measurements (N), and the resulting uniform disk angular diameters are listed together with the reduced $\chi^2$.  The baseline used is also indicated. The pulsation phases are calculated using the ephemeris from \citet{kukarkin71}: $T_\mathrm{0}=2427628.86$ days and $P=5.3663$ days. There is no error on the phase at 0.300 since the two measurements are from the same date but at different reference wavelengths. The S1S2 data were merged in order to derive a mean uniform disk angular diameter.}
\begin{center}
\begin{tabular}{|ll|ll|c|}
\hline

\hline
 $\phi$                                                          & N & $\theta_\mathrm{UD}$                                           & $\chi^{2}_\mathrm{red}$  & baseline   \\
                                                                         &       &                  [mas]                                          &                                                    & \\
\hline
                $0.300  $                                               & 2                       &       \multirow{4}{*}{1.705$_\mathrm{\pm       0.035}$}                   &  \multirow{4}{*}{0.9}    & S1S2\\
                0.543$_\mathrm{\pm         0.009        }$      & 4                      &                                                                                                         &                                      & S1S2 \\
                0.666$_\mathrm{\pm      0.006   }$      & 3                     &                                                                                                        &                                       &       S1S2\\
                0.821$_\mathrm{\pm      0.012   }$      &  3                    &                                                                                                       &                                &       S1S2\\
\hline
        0.180$_\mathrm{\pm      0.006   }$      & 15 &  1.418$_\mathrm{\pm      0.027   }$&     1.2                      &  E1E2 \\
        0.192$_\mathrm{\pm      0.015   }$      & 7   & 1.400$_\mathrm{\pm      0.010   }$&     0.1                               & E1E2\\
        0.292$_\mathrm{\pm      0.009   }$      & 9                     &       1.454$_\mathrm{\pm       0.022   }$&     1.6                     & E1E2\\
         0.401$_\mathrm{\pm         0.003         }$     & 4                       &      1.439$_\mathrm{\pm      0.026  }$  & 4.4               &  E1E2   \\
        0.526$_\mathrm{\pm      0.001   }$      & 12                    &       1.425$_\mathrm{\pm       0.012   }$&     0.6     &               E1E2\\
        0.627$_\mathrm{\pm      0.017   }$      & 7                     &       1.436$_\mathrm{\pm       0.019   }$&     0.6                             & E1E2 \\
        0.741$_\mathrm{\pm      0.003   }$      & 8                      &       1.318$_\mathrm{\pm      0.017    }$&    2.4                             & E1E2 \\
        0.783$_\mathrm{\pm      0.002   }$      & 15                     &       1.316$_\mathrm{\pm       0.010  }$&     0.9                             & E1E2 \\
        0.821$_\mathrm{\pm      0.012   }$      & 6                     &       1.393$_\mathrm{\pm       0.025   }$&     2.2                             & E1E2\\
        0.848$_\mathrm{\pm      0.007   }$      & 24                     &       1.360$_\mathrm{\pm       0.020  }$&     3.3                             &E1E2 \\
        0.893$_\mathrm{\pm      0.005   }$      & 30                    &       1.405$_\mathrm{\pm       0.019   }$&     8.1                     &E1E2\\
        0.927$_\mathrm{\pm      0.003   }$      & 8 &   1.353$_\mathrm{\pm      0.029   }$&     0.9                      & E1E2\\
        0.942$_\mathrm{\pm      0.007   }$      & 13 &  1.408$_\mathrm{\pm      0.026   }$&     1.1                          & E1E2\\
        0.999$_\mathrm{\pm      0.025   }$      & 8                     &       1.337$_\mathrm{\pm       0.016   }$&     1.4             &E1E2 \\
\hline
\end{tabular}
\end{center}
\end{table}

\begin{table}[htbp]
\caption{\label{Tab.spips} Angular diameters as derived from the application of the SPIPS algorithm to $\delta$~Cep, using photometry, interferometry, and the cross-correlated radial velocity curves as inputs (see \citet{merand15}): $\theta_{\mathrm{Ross}}$ is the Rosseland angular diameter (provided as an indication, but not used in this paper); $\theta_{\mathrm{UD}}$[0.8~$\mu$m] and $\theta_{\mathrm{UD}}$[2.2~$\mu$m]  are the UD angular diameters calculated at 0.8 and 2.2~$\mu$m, respectively. We used a distance $d$ of $274\pm11$~pc \citep{benedict02}.}
\begin{center}
\begin{tabular}{lccc}
\hline
\hline
phase  & $\theta_{\mathrm{Ross}}$ & $\theta_{\mathrm{UD}}$[0.8$\mu$m] & $\theta_{\mathrm{UD}}$[2.2$\mu$m]  \\
\hline
0.00    &       1.399   &       1.337   &       1.385   \\
0.02    &       1.409   &       1.347   &       1.396   \\
0.04    &       1.420   &       1.357   &       1.406   \\
0.06    &       1.430   &       1.366   &       1.415   \\
0.08    &       1.439   &       1.373   &       1.423   \\
0.10    &       1.449   &       1.382   &       1.432   \\
0.12    &       1.457   &       1.390   &       1.440   \\
0.14    &       1.465   &       1.398   &       1.448   \\
0.16    &       1.473   &       1.405   &       1.455   \\
0.18    &       1.480   &       1.412   &       1.462   \\
0.20    &       1.487   &       1.416   &       1.467   \\
0.22    &       1.492   &       1.421   &       1.472   \\
0.24    &       1.498   &       1.426   &       1.477   \\
0.26    &       1.502   &       1.431   &       1.482   \\
0.28    &       1.507   &       1.435   &       1.486   \\
0.30    &       1.510   &       1.438   &       1.489   \\
0.32    &       1.513   &       1.441   &       1.492   \\
0.34    &       1.515   &       1.443   &       1.494   \\
0.36    &       1.517   &       1.444   &       1.496   \\
0.38    &       1.518   &       1.442   &       1.495   \\
0.40    &       1.518   &       1.442   &       1.495   \\
0.42    &       1.518   &       1.442   &       1.495   \\
0.44    &       1.517   &       1.441   &       1.494   \\
0.46    &       1.515   &       1.440   &       1.493   \\
0.48    &       1.513   &       1.438   &       1.491   \\
0.50    &       1.511   &       1.435   &       1.488   \\
0.52    &       1.508   &       1.432   &       1.485   \\
0.54    &       1.504   &       1.429   &       1.481   \\
0.56    &       1.499   &       1.424   &       1.477   \\
0.58    &       1.494   &       1.420   &       1.472   \\
0.60    &       1.489   &       1.414   &       1.467   \\
0.62    &       1.482   &       1.408   &       1.461   \\
0.64    &       1.476   &       1.402   &       1.454   \\
0.66    &       1.468   &       1.395   &       1.447   \\
0.68    &       1.460   &       1.387   &       1.440   \\
0.70    &       1.452   &       1.379   &       1.432   \\
0.72    &       1.443   &       1.371   &       1.423   \\
0.74    &       1.433   &       1.362   &       1.414   \\
0.76    &       1.423   &       1.352   &       1.404   \\
0.78    &       1.412   &       1.342   &       1.394   \\
0.80    &       1.401   &       1.331   &       1.383   \\
0.82    &       1.390   &       1.320   &       1.372   \\
0.84    &       1.379   &       1.313   &       1.363   \\
0.86    &       1.370   &       1.305   &       1.355   \\
0.88    &       1.364   &       1.302   &       1.351   \\
0.90    &       1.362   &       1.300   &       1.349   \\
0.92    &       1.365   &       1.304   &       1.353   \\
0.94    &       1.371   &       1.310   &       1.358   \\
0.96    &       1.379   &       1.318   &       1.366   \\
0.98    &       1.389   &       1.327   &       1.376   \\
\hline
          &   mas   & mas & mas \\
\hline
\end{tabular}
\end{center}
\end{table}


\section{ Uniform disk (UD) angular diameters}\label{s_UD}

In Figs.~\ref{Fig.all1} and \ref{Fig.all2}, the calibrated visibilities corresponding to E1E2 measurements are plotted as a function of the spatial frequency together with the corresponding (u,v) coverage and for each pulsation phase. We fit these calibrated visibilities by a  uniform disk using a JMMC\footnote{http://www.jmmc.fr/litpro} tool, {\it LITpro} \citep{tallonbosc08}. The results are given in Tab.~\ref{Tab.ang} and plotted in Fig.~\ref{Fig.diam} with blue crosses. The data corresponding to S1S2 measurements are shown in Fig.~\ref{Fig.S1S2_CSE}a. Owing to their low spatial frequencies, these S1S2 measurements are almost not sensitive to the pulsation of the star. This can be seen in  Fig.~\ref{Fig.S1S2_CSE}b. Moreover, we have only a few S1S2 measurements per pulsation phase (four at most). Consequently, as a point of comparison with the E1E2 data, we merge the S1S2 data and fit them with a uniform disk angular diameter. The result is given in Tab.~\ref{Tab.ang} and plotted in Fig.~\ref{Fig.diam} by a horizontal dashed zone. For the data of 2014 October 24, which include both E1E2 and S1S2 baselines, we made two fits, one including all the S1S2 data and another with the E1E2 data. 
We overplot the very precise K-band uniform disk angular diameter curve obtained by \citet{merand05} (magenta dots).
To analyse our VEGA/CHARA angular diameter measurements, we consider the Spectroscopy-Photometry-Interferometry for Pulsating Stars algorithm (SPIPS; \citet{merand15}). The SPIPS code combines all the available observables of $\delta$~Cep: radial velocimetry \citep{bersier94b, storm04}, interferometry (FLUOR/CHARA data), and photometry in the V \citep{berdnikov02,engle14,kiss98,moffett84}, J, H, and K bands  \citep{barnes97} in order to estimate  
the projection factor, the variation of the effective temperature, and the H and K band excesses.  The data used for the fit for $\delta$~Cep have been published in digital form\footnote{http://vizier.cfa.harvard.edu/viz-bin/VizieR?-source=J/A+A/584/A80}. In the list of these outputs, we are particularly interested in the uniform disk angular diameter curves, calculated at 2200~nm and 800~nm, and derived directly from the framework of \citet{merand15}. The data are given in Tab.~\ref{Tab.spips} and are shown in Fig.~\ref{Fig.diam} by red crosses and light blue squares, respectively. In the infrared, the K-band UD angular diameter curve of SPIPS is consistent with the FLUOR/CHARA measurements. In the visible domain, the reduced $\chi^2$ between the VEGA~(E1E2) and SPIPS UD angular diameters is  9, while it increases slightly to 11 when we replace the SPIPS angular diameter variation by a constant corresponding to the average of the VEGA UD angular diameters, i.e.  $1.376\pm0.030$ mas. We thus detect a pulsation in the visible band, but   make three remarks:

\begin{figure}[htbp]
\begin{center}
\resizebox{0.9\hsize}{!}{\includegraphics[clip=true]{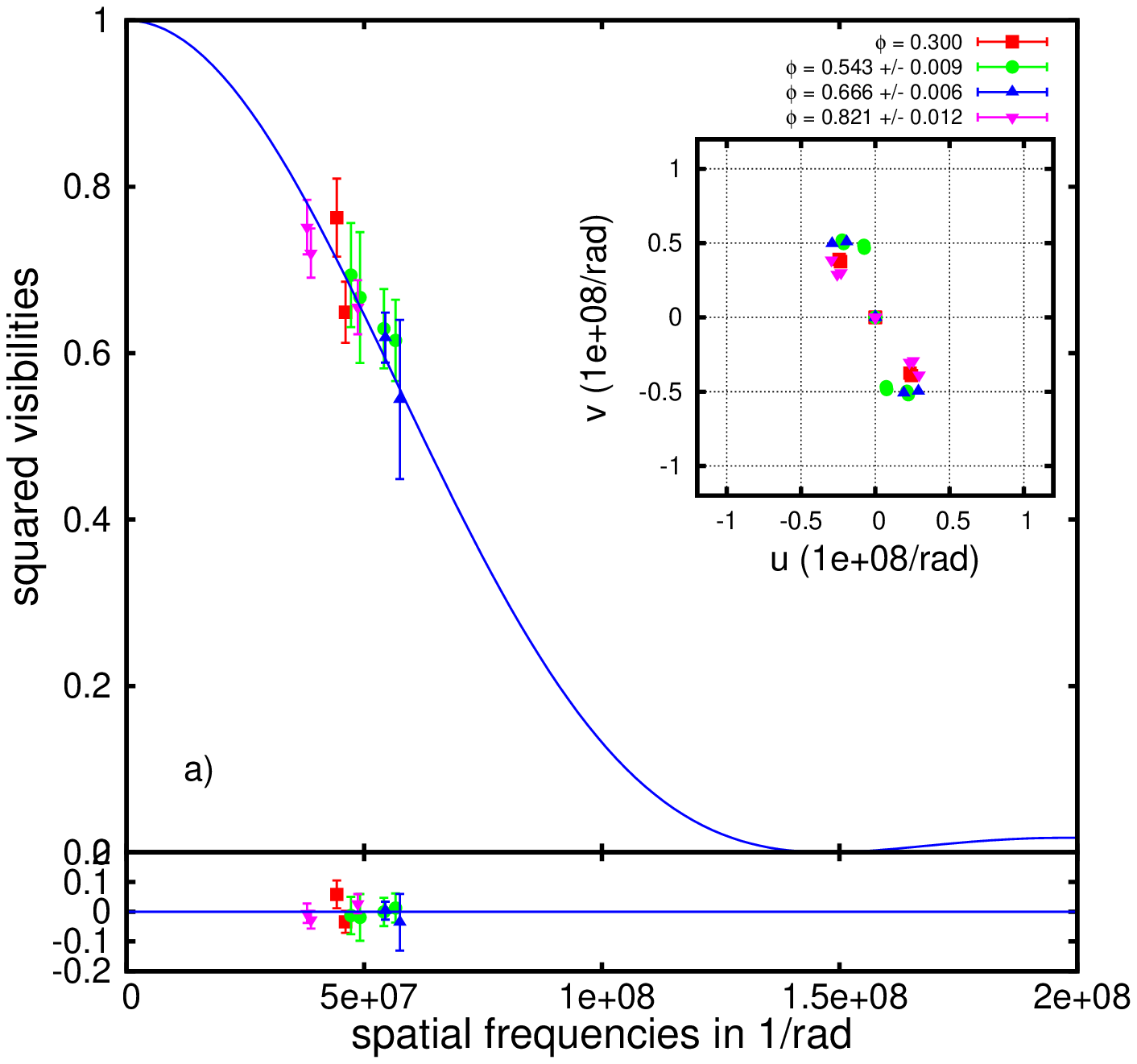}}
\resizebox{0.9\hsize}{!}{\includegraphics[clip=true]{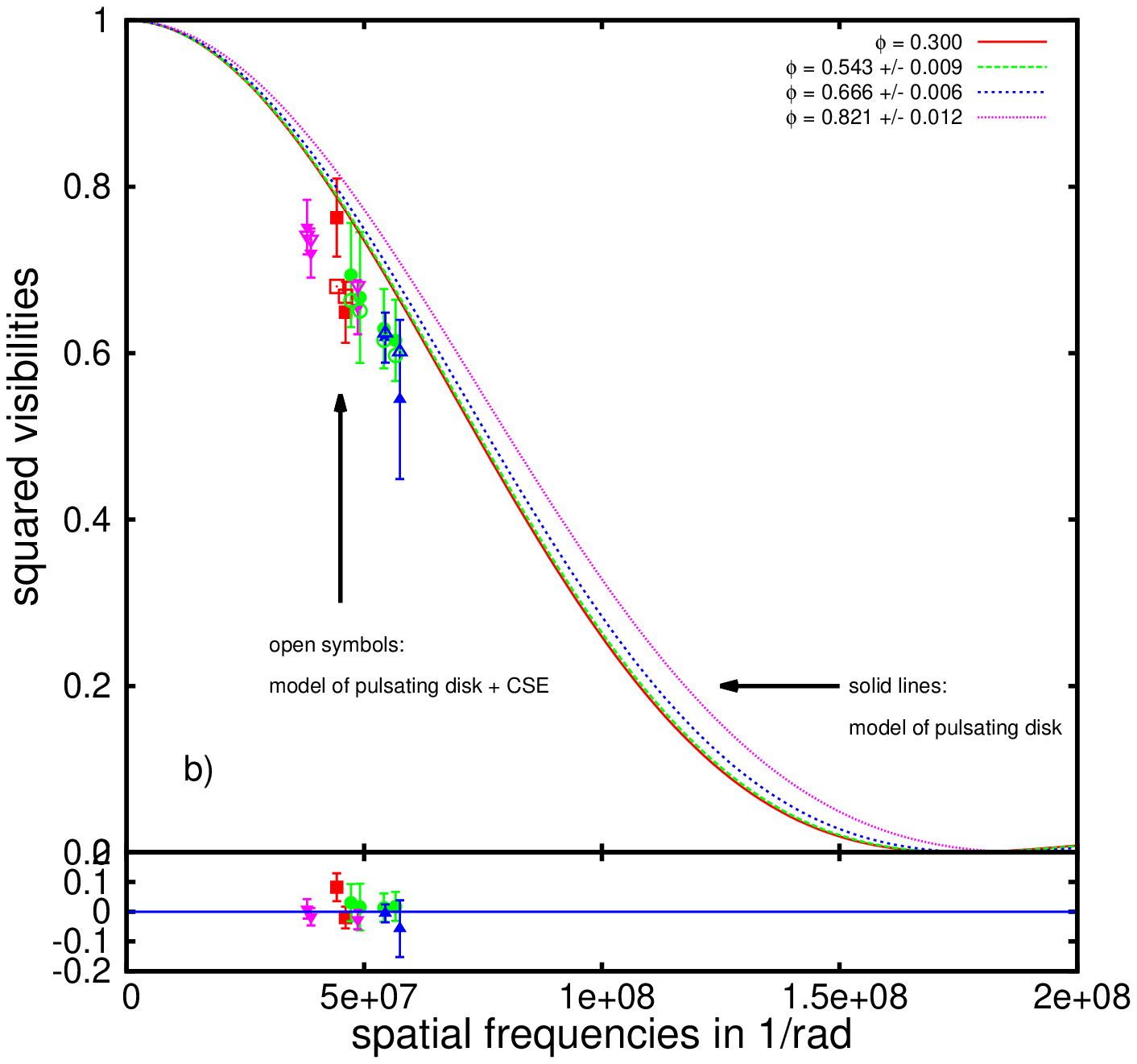}}
\caption{a)  S1S2 observed calibrated squared visibilities secured at four different pulsation phases  plotted as a function of the spatial frequency together with the best fit of uniform disk model (solid blue line). In the upper right corner we show the corresponding (u,v) coverage. b)  Same data  fitted with a two-component model (open symbols) including a uniform pulsating disk and a CSE (see  text). As a comparison, the solid lines are the corresponding one-component models, i.e. a uniform pulsating disk (in the R band) without the CSE.} \label{Fig.S1S2_CSE}
\end{center}
\end{figure}


\begin{enumerate}
\item The VEGA S1S2 measurements provide a mean $\theta_\mathrm{UD}$ (with a reduced $\chi^2$ of 0.9), which is significantly larger than the E1E2 angular diameter curve by about 10$\sigma$. This suggests the presence of a large angular structure around the pulsating Cepheid. 
\item The SPIPS predictions are consistent with the VEGA $\theta_\mathrm{UD}$ angular diameters from 0.0 to 0.8 in phase (with a reduced $\chi^2$ of 4).  This basically means that $\delta$~Cep is pulsating in a quasi-hydrostatic way over this range of pulsation phase, even if the two measurements (at $\phi= 0.741$ and $\phi=0.783$) are slightly below the SPIPS curve (around 2$\sigma$). 
\item Five measurements between 0.8 and 1.0 are clearly from 2 to 6$\sigma$ above the SPIPS curve (with a reduced $\chi^2$ of 8). 
\end{enumerate}

\section{Detection of a circumstellar environment or a nebulae in the visible}\label{s_CSE}


In Fig.~\ref{Fig.S1S2_CSE}b, the S1S2 data are fitted with a model (open symbols) composed of the UD derived from the SPIPS UD angular diameter curve at the corresponding phases of VEGA observations and a second static disk describing the CSE. The orientation of the S1S2  measurements in the (u,v) coverage are between 148 and  171 degrees,  preventing us from investigating a non-centrosymmetric CSE. The best-fit parameters are $\theta_\mathrm{CSE}=8.9\pm3.0$~mas (the size of the CSE) and  $ \frac{f_\mathrm{cse}}{f_\mathrm{\star}}=0.07\pm0.01$ (the relative flux contribution of the CSE, where $f_\mathrm{cse}$ and  $f_\mathrm{\star}$ are the fluxes of the CSE and the star in the R band, respectively) with a reduced $\chi^2$ of 0.6. The two parameters,  however, are significantly correlated (0.8) owing to the low number of S1S2 observations. Interestingly, we can also fit the S1S2 data (with the same reduced $\chi^2$ of 0.6) using a model composed of a pulsating disk, as in the previous case, together with a background  contribution of $0.07\pm0.01$ in flux filling the field of view of the interferometer.
This model physically corresponds to the visible counterpart of the infrared nebulae discovered by \citet{marengo10} and the HI nebula found with the Very Large Array (VLA) \citep{matthews12}. Unfortunately, our S1S2 data are  too scarce to distinguish between these two hypotheses.

We have fitted the S1S2 data with a two-component model without considering the E1E2 data. Hence, we now have  to verify whether this two-component model is consistent with our E1E2 data, in particular at minimum radius.  In Sect.~3, we  find that the E1E2 measurements are not consistent with a quasi-hydrostatic pulsating disk at the minimum radius of the star (between phase 0.8 and 1.0).

\begin{figure*}[htbp]
\begin{center}
\resizebox{0.95\hsize}{!}{\includegraphics[clip=true]{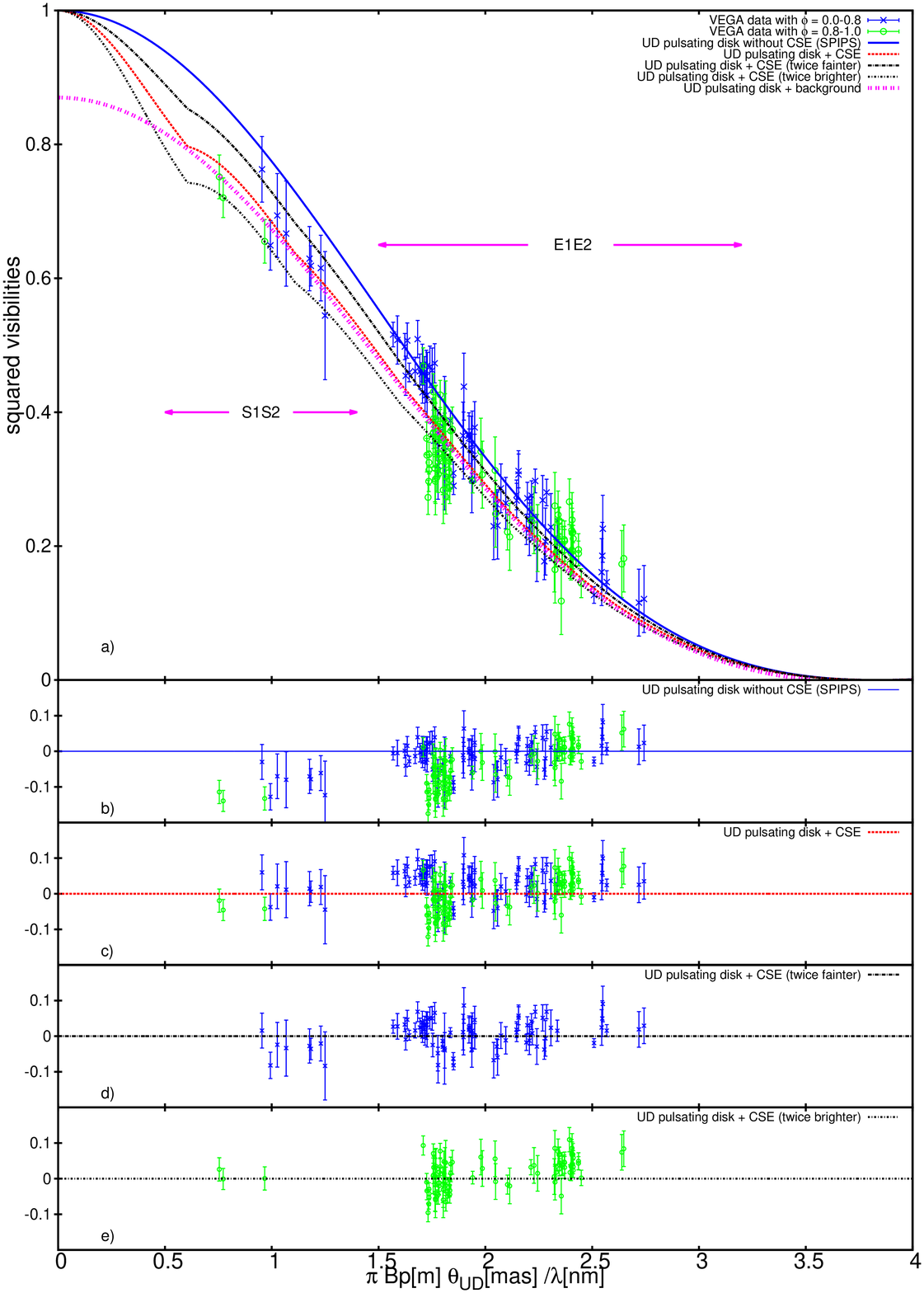}}
\end{center}
\vspace{-1.8cm}
\caption{In panel a), the VEGA squared visibilities are plotted as a function of $x = \frac{\pi B_p \mathrm{[m]} \theta_\mathrm{UD} \mathrm{[mas]} }{\lambda \mathrm{[nm]}}$ in order to allow a comparison between data of different phases (i.e. different $\theta_\mathrm{UD}$) and effective wavelengths. Five models are overplotted: (1) the UD pulsating disk without CSE from the SPIPS algorithm (blue solid line), (2) the SPIPS UD pulsating disk + the CSE from Eq. 1 (red dotted line),  (3) the SPIPS UD pulsating disk + the CSE from Eq. 1 but two times brighter  (dashed line),  (4) the SPIPS UD pulsating disk + the CSE or Eq. 1 but two times fainter (dash-dotted line), (5) the SPIPS UD pulsating disk + a background contribution filling the field of view of the interferometer (magenta dotted line). The residual between the observations and  models 1 to 4 are shown in panels, b), c), d), and e), respectively. In panels d) and e) we show only the data corresponding to phases intervals $\phi=0.0-0.8$ and $\phi=0.8-1.0$, respectively.}  \label{Fig.V2}
\end{figure*}

In order to address this issue, we plot in panel~a) of Fig.~\ref{Fig.V2} all the VEGA visibilities measured as a function of $x=\frac{\pi B_p \mathrm{[m]} \theta_\mathrm{UD} \mathrm{[mas]}}{\lambda \mathrm{[nm]}}$, where $\theta_\mathrm{UD}$ is fixed and interpolated from the SPIPS UD angular diameter curve at the corresponding pulsation phase of VEGA observations. The data are rescaled in such a way that they can be compared, despite different $\theta_\mathrm{UD}$ (or pulsation phases) and different wavelengths of observation. This is possible since the reference is the SPIPS $\theta_\mathrm{UD}$ semi-theoretical curve represented by the solid blue line expressed by $V^2(x)= | \frac{2 J_1(x)}{x} |$, where $J_1$ is the Bessel function of the first order. This concept of pseudo-baseline was first introduced by \citet{merand06}. Using this approach, we confirm the three statements found in Sect.~3:

\begin{enumerate}
\item  The S1S2 measurements, whatever the pulsation phase, are significantly lower than in the rescaled uniform disk model (see the S1S2 measurements, i.e. data with $x$ from approximately 0.5 to 1.4, Fig.~\ref{Fig.V2}b). This deviation is removed as soon as we consider a resolved structure around $\delta$~Cep (see below).
\item The data with a pulsating phase from $\phi=0.0$ to $\phi=0.8$ are consistent with the uniform disk model for E1E2 measurements (blue crosses with $x$ from 1.5 to 2.8 in Fig.~\ref{Fig.V2}b).
\item The data with a pulsating phase from $\phi=0.8$ to $\phi=1.0$ are not consistent with the uniform disk model for E1E2 measurements (green circles around $x=1.7$ in Fig.~\ref{Fig.V2}b).
\end{enumerate}

In Fig. \ref{Fig.V2}a, we overplot a red dotted line to the squared visibility curve corresponding to a uniform disk surrounded by the CSE using the  formula

\begin{equation}
V^2(x)=(f_{\star} | \frac{2 J_1(x)}{x} |+f_{\mathrm{cse}} | \frac{2 J_1(\mathrm{s_r} x)}{\mathrm{s_r} x} | )^2
,\end{equation}

where $f_\mathrm{cse} = 0.065$ and $f_\mathrm{\star} = 0.935$ (corresponding to $ \frac{f_\mathrm{cse}}{f_\mathrm{\star}}=0.07\pm0.01$ and $f_\mathrm{cse} + f_\mathrm{\star} = 1$), $\mathrm{s_r}$ is the mean size ratio between the CSE and the pulsating disk. Considering the mean UD angular diameter of the SPIPS curve ($\theta_\mathrm{mean}=1.39$ mas), we find  $\mathrm{s_r} = 6.32$.  The result is unchanged (i.e. within the width of the line) if we use the 6.75 and 6.07 values corresponding to the minimum and maximum UD angular diameters.

We find that the S1S2 measurements are properly fitted by this two-component model, as expected (Fig.~\ref{Fig.V2}c). However, around $x=1.7$ the E1E2 measurements corresponding to $\phi=0.0-0.8$ and $\phi=0.8-1.0$ are respectively above and below the curve (by about 2$\sigma$). All the E1E2 measurements are partly above the curve (about $1\sigma$) at larger frequencies ($x > 2.3 $).

If we consider the model composed of a uniform disk surrounded by a background, we obtain
\begin{equation}
V^2(x)=(f_{\star} | \frac{2 J_1(x)}{x} |)^2-f_{\mathrm{cse}}^2,
\end{equation}
and the corresponding curve is plotted by a  magenta dotted line in Fig.~\ref{Fig.V2}a. We arrive at  the same conclusions as the model of the pulsating uniform disk surrounded by a CSE. In particular, we find the presence of two regimes in phase, {even} considering a CSE or a background. We explore two hypotheses.

First, we cannot fit all the VEGA measurements at the same time if we consider a CSE two times fainter or brighter (see black dotted and dot-dashed lines in Fig. \ref{Fig.V2}a).
Qualitatively, however,  it seems that considering a CSE two times fainter between phase 0.0 and 0.8 is a good compromise to fit the S1S2 and E1E2 measurements together, even if not totally satisfactory (see Fig. \ref{Fig.V2}d). To consider a CSE two times brighter between phase 0.8 and 1.0 (when the size of the star is at a  minimum, and is hot and bright) helps to fit the data except at large spatial frequencies (see Fig. \ref{Fig.V2}e). If this hypothesis is correct, then the  $\delta$~Cep would light up its environment (CSE or background) differently at maximum and minimum radius. 

The second possibility is to consider that the CSE (or the background) has a constant brightness and that the discrepancy found for E1E2 measurements comes mainly from an additional structure angularly smaller ($\phi=0.0-0.8$) or larger ($\phi=0.8-1.0$) in size than the SPIPS $\theta_\mathrm{UD}$ stellar pulsating disk,  for instance a strong limb-darkening variation. The limb-darkening coefficient, defined as $ k=\frac{\theta_{\mathrm{UD}}}{\theta_{\mathrm{LD}}}$, is used to convert the $\theta_\mathrm{UD}$ diameters into limb-darkened $\theta_\mathrm{LD}$ angular diameters. The $k$ coefficient variation in the visible band (at 600~nm) and over the cycle of pulsation of the star is found to be from the hydrodynamical model of 0.015 \citep{nardetto06b}, which corresponds to about 1\%.  Similarly, \citet{marengo02, marengo03} have found  a variation for the k-parameter of about 0.01 and 0.02 for $\zeta$~Gem and at 570nm using quasi-hydrostatic and hydrodynamical models, respectively. Also for comparison, \citet{davis09} used a phase-dependent k-parameter to analyse their SUSI data of $\ell$~Car, and found a variation of $0.012$. 
This means typically a 1\% difference in $\theta_\mathrm{UD}$, and thus a 1\% difference in $x$, which corresponds to 0.02 in absolute value around $x=1.7$ in Fig. \ref{Fig.V2}. Consequently, the limb-darkening variation of $\delta$~Cep is one order of magnitude lower than the discrepancies found for the E1E2 measurements. In the case of $\delta$~Cep, the theoretical average value of $k$ is of $0.954$ \citep{nardetto06b}. So far, the only published direct measurement of the limb-darkening coefficient of a Cepheid (in the visible) has been performed by \citet{pilecki13}  for a Cepheid ($P$=3.80~d) in an eclipsing binary system in the LMC. They find the limb-darkening effect to be much stronger than expected for a star of that temperature, giving a possible range for $k$ from 0.91 to 0.93, assuming however no temperature dependence. Recently, \citet{gieren15} studied another Cepheid in the LMC ($P$=2.99~d) and found a similar rather strong limb-darkening effect. If we consider the lower $k$ value found by \citet{pilecki13}, all the E1E2 measurements in Fig. \ref{Fig.V2} should translate towards lower values of $x$ --  4\% or about 0.06. The measurements in blue crosses in the figure would be better fitted, while green open circles would not (in particular near $x \simeq 1.7$). The limb-darkening effect is probably not the key issue to explain the discrepancy found for the E1E2 measurements.

\section{Discussion}\label{s_C}


We observed   $\delta$~Cep intensively on twenty nights. The initial purpose of these observations was to apply the BW method, but instead we detected for the very first time a static resolved structure around $\delta$~Cep contributing to about 7\% of  the total flux in the visible.  This is not totally surprising as  envelopes around Cepheids have already been discovered by long-baseline interferometry in the K band with VLTI and CHARA \citep{kervella06a, merand06}. In addition, four Cepheids have also been observed in the N band with VISIR and MIDI \citep{kervella09,gallenne13b} and one with NACO \citep{gallenne11,gallenne12}. Some evidence has also been found using high-resolution spectroscopy \citep{nardetto08b}.  From these observations, the typical size of the envelope of Cepheids seems to be around 3 stellar radii and the flux contribution from 2\% to 10\% of the continuum in the K band for medium- and long-period Cepheids, respectively, while it is around 10\% or more in the N band. However, \citet{merand05} do not mention any contribution from a CSE in the case of $\delta$~Cep in their infrared FLUOR data, while \citet{merand06} found an improved agreement with a larger set of data when considering a model with a CSE. They used a ring model   $3.54$ mas in size,  0.5 mas in width, and with a relative brightness of 1.5\%, which is confirmed in \citet{merand15}. The processes at work in infrared and in the visible regarding the CSE are different. We expect thermal emission in the infrared and scattering in the visible.  It is also worth mentioning that the mid-infrared flux in excess is not necessarily the evidence for mass loss (see e.g.  \citealt{schmidt15}) and that the resolved structure we observe around $\delta$~Cep, instead of a CSE, could be simply a visible nebulae (also contributing  to 7\% of the total flux), as reported in the infrared by \citet{marengo10} and in the radio domain by \citet{matthews12}.

Our second result is the presence of an additional second-order discrepancy between the observations and the models (pulsating disk + CSE or pulsating disk + background) at high spatial frequencies, which is not clearly understood. One possibility could be that the star is lighting up its environment differently at minimum ($\phi \simeq 0.8-1.0$) and maximum ($\phi \simeq 0.0-0.8$) radius. This reverberation effect would then be   more important in the visible than  in the infrared since the contribution in flux of the CSE (or the background) in the visible band is  about 7\% compared to 1.5\% in the infrared. Interestingly, the phase interval from 0.8 to 1.0 is usually disregarded in the infrared surface brightness version of the BW method owing  to general poor agreement between spectroscopic and photometric angular diameters \citep{storm11a,storm11b}. The other possibility is to consider a {\it static} environment with a constant brightness in time, but with an additional dynamical effect. However, from the hydrodynamical simulations and observations of Cepheids in eclipsing binary systems, it seems that the limb-darkening effect in the visible band does not vary strongly enough to reproduce the second-order discrepancy found in our interferometric observations. Also possible is a strong compression or shockwave occurring near the photosphere at the minimum radius, which is indeed the layer probed by photometry and interferometry. However, such a shock is not seen in the hydrodynamical code \citep{nardetto04} or in the spectroscopic data, conversely to X~Sgr, an atypical Cepheid in which a shockwave has  indeed been reported \citep{mathias06}. Why this dynamical effect is seen in the visible band and not in the infrared is another pending question in this hypothesis. We also exclude non-radial pulsation, since very accurate space photometry does not support the detection of such modes in Galactic classical Cepheids  (e.g. \citealt{poretti15}), and the previous spectroscopic campaigns would have shown up quite easily. Furthermore, \citet{anderson15a} found the signature of a companion in recent spectroscopic data of $\delta$~Cep. The expected flux contribution is supposed to be lower than 1\% in the visible, clearly not detectable by VEGA/CHARA.  Finally, reverberation seems to be the more plausible hypothesis. More interferometric data on Cepheids in the visible might shed light on it. In particular, covering  the spatial frequencies properly ($x$ from 0.2 to 3) at a unique pulsating phase (maximum and minimum radii) is probably required to go further in the analysis. 


\begin{acknowledgements}
The authors acknowledge the support of the French Agence Nationale de la Recherche (ANR), under grant ANR-15-CE31-0012- 01 (project UnlockCepheids). We acknowledge financial support from ``Programme National de Physique Stellaire'' (PNPS) of CNRS/INSU, France. The CHARA Array is funded by the National Science Foundation through NSF grants AST-0606958 and AST-0908253 and by Georgia State University through the College of Arts and Sciences, as well as the W. M. Keck Foundation. WG gratefully acknowledges financial support for this work from the BASAL Centro de Astrofisica y Tecnologias Afines (CATA) PFB-06/2007, and from the Millenium Institute of Astrophysics (MAS) of the Iniciativa Cientifica Milenio del Ministerio de Economia, Fomento y Turismo de Chile, project IC120009. We acknowledge financial support for this work from ECOS-CONICYT grant C13U01. Support from the Polish National Science Center grant MAESTRO 2012/06/A/ST9/00269 is also acknowledged. EP and MR acknowledge financial support from PRIN INAF-2014. NN, PK, AG, and WG acknowledge support from the French-Chilean exchange program ECOS- Sud/CONICYT (C13U01). This project was partially supported by the Polish Ministry of Science grant Ideas Plus. This research has made use of the SIMBAD and VIZIER\footnote{Available at http://cdsweb.u- strasbg.fr/} databases at CDS, Strasbourg (France),   the Jean-Marie Mariotti Center \texttt{Aspro} service\footnote{Available at http://www.jmmc.fr/aspro}, and the electronic bibliography maintained by the NASA/ADS system. This research has made use of the Jean-Marie Mariotti Center \texttt{SearchCal} service \footnote{Available at http://www.jmmc.fr/searchcal} co-developed by FIZEAU and LAOG/IPAG, and  CDS Astronomical Databases SIMBAD and VIZIER \footnote{Available at http://cdsweb.u-strasbg.fr/}. This research has made use of the Jean-Marie Mariotti Center \texttt{LITpro} service co-developed by CRAL, IPAG, and LAGRANGE. \footnote{LITpro software available at http://www.jmmc.fr/litpro}. This research has made use of the Jean-Marie Mariotti Center \texttt{OiDB} service \footnote{Available at http://oidb.jmmc.fr }. 
\end{acknowledgements}

\bibliographystyle{aa}  
\bibliography{bibtex_nn} 


\begin{appendix}

\section{VEGA observations: Log, transfer functions, and visibility curves}
\begin{table*}
\begin{center}
\caption[]{Observing log with the E1E2 baseline (part 1). The columns give, respectively, the date, the RJD, the hour angle (HA), the minimum and maximum wavelengths over which the squared visibility is calculated, the projected baseline length Bp and its orientation PA, the signal-to-noise ratio on the fringe peak; the last column provides the calibrated squared visibility $V^{2}$ together with the statistic error on $V^{2}$, and the systematic error on $V^{2}$ (see text for details). The data are available on the Jean-Marie Mariotti Center \texttt{OiDB} service (Available at http://oidb.jmmc.fr).}
\label{Tab.log1}
\setlength{\doublerulesep}{\arrayrulewidth}
\begin{tabular}{lcccccccccc}
\hline \noalign{\smallskip}
\hline\hline\hline\hline
  &    Date             &   RJD    & HA  &  $\lambda_\mathrm{min}$ & $\lambda_\mathrm{max}$ & Bp &  PA    &   S/N &$V^{2}_\mathrm{cal \pm stat \pm syst}$ \\
                       & [ yyyy.mm.dd ]&  [ days ] & [ hour ] &             [ nm ]&                     [ nm ]  &    [ m ] & [ deg ] &          &                                                         \\
                       \hline                                                                                                                                   
$\phi=0.180\pm0.006$
&       2013.10.26      &       56591.755       &       2.19    &       535     &       555     &       65.85   &        -149.17 &       5       &       0.226   $_\mathrm{\pm   0.050   \pm     0.003   }$\\
&       2013.10.26      &       56591.755       &       2.19    &       710     &       730     &       65.85   &        -149.16 &       7       &       0.352   $_\mathrm{\pm   0.050   \pm     0.003   }$\\
&       2013.10.26      &       56591.755       &       2.19    &       725     &       745     &       65.85   &        -149.16 &       9       &       0.438   $_\mathrm{\pm   0.050   \pm     0.003   }$\\
&       2013.10.26      &       56591.770       &       2.55    &       535     &       555     &       65.79   &        -153.92 &       4       &       0.186   $_\mathrm{\pm   0.050   \pm     0.003   }$\\
&       2013.10.26      &       56591.770       &       2.55    &       710     &       730     &       65.79   &        -153.91 &       6       &       0.300   $_\mathrm{\pm   0.050   \pm     0.002   }$\\
&       2013.10.26      &       56591.770       &       2.55    &       725     &       745     &       65.79   &        -153.91 &       7       &       0.350   $_\mathrm{\pm   0.050   \pm     0.003   }$\\
&       2013.10.26      &       56591.785       &       2.91    &       535     &       555     &       65.72   &        -158.70 &       3       &       0.161   $_\mathrm{\pm   0.050   \pm     0.002   }$\\
&       2013.10.26      &       56591.785       &       2.91    &       710     &       730     &       65.72   &        -158.70 &       7       &       0.344   $_\mathrm{\pm   0.050   \pm     0.003   }$\\
&       2013.10.26      &       56591.785       &       2.91    &       725     &       745     &       65.72   &        -158.70 &       7       &       0.365   $_\mathrm{\pm   0.050   \pm     0.003   }$\\
&       2014.07.05      &       56843.917       &       -1.27   &       730     &       750     &       62.08   &        -106.85 &       6       &       0.320   $_\mathrm{\pm   0.050   \pm     0.002   }$\\
&       2014.07.05      &       56843.933       &       -0.87   &       730     &       750     &       63.14   &        -111.65 &       4       &       0.354   $_\mathrm{\pm   0.050   \pm     0.002   }$\\ 
&       2014.07.05      &       56843.958       &       -0.17   &       490     &       510     &       64.52   &        -120.01 &       2       &       0.116   $_\mathrm{\pm   0.050   \pm     0.002   }$\\
&       2014.07.05      &       56843.958       &       -0.16   &       660     &       680     &       64.53   &        -120.02 &       5       &       0.230   $_\mathrm{\pm   0.050   \pm     0.002   }$\\
&       2014.07.05      &       56843.979       &       0.26    &       490     &       510     &       65.09   &        -125.08 &       2       &       0.121   $_\mathrm{\pm   0.050   \pm     0.002   }$\\
&       2014.07.05      &       56843.979       &       0.26    &       660     &       680     &       65.10   &        -125.09 &       5       &       0.231   $_\mathrm{\pm   0.050   \pm     0.002   }$\\
\hline
$\phi=0.192\pm0.015$
&       2012.09.24      &       56194.645       &       -2.49   &       547     &       560     &       57.61   &        -91.87  &       7       &       0.241   $_\mathrm{\pm   0.050   \pm     0.002   }$\\
&       2012.09.24      &       56194.645       &       -2.51   &       710     &       730     &       57.53   &        -91.65  &       28      &       0.457   $_\mathrm{\pm   0.050   \pm     0.002   }$\\ 
&       2012.09.24      &       56194.645       &       -2.51   &       730     &       750     &       57.53   &        -91.65  &       25      &       0.498   $_\mathrm{\pm   0.050   \pm     0.003   }$\\ 
&       2012.09.24      &       56194.760       &       0.26    &       532     &       547     &       65.10   &        -125.10 &       9       &       0.146   $_\mathrm{\pm   0.050   \pm     0.002   }$\\
&       2012.09.24      &       56194.760       &       0.26    &       547     &       560     &       65.10   &        -125.10 &       10      &       0.127   $_\mathrm{\pm   0.050   \pm     0.002   }$\\
&       2012.09.24      &       56194.760       &       0.26    &       710     &       730     &       65.09   &        -125.06 &       39      &       0.366   $_\mathrm{\pm   0.050   \pm     0.003   }$\\ 
&       2012.09.24      &       56194.760       &       0.26    &       730     &       750     &       65.09   &        -125.06 &       28      &       0.387   $_\mathrm{\pm   0.050   \pm     0.003   }$\\ 
\hline
$\phi=0.292\pm0.009$
&       2014.07.11      &       56849.865       &       -2.14   &       585     &       600     &       59.09   &        -96.30  &       12      &       0.312   $_\mathrm{\pm   0.025   \pm     0.003   }$\\
&       2014.07.11      &       56849.865       &       -2.14   &       730     &       750     &       59.08   &        -96.26  &       12      &       0.454   $_\mathrm{\pm   0.039   \pm     0.002   }$\\
&       2014.07.11      &       56849.876       &       -1.87   &       585     &       600     &       60.10   &        -99.54  &       10      &       0.225   $_\mathrm{\pm   0.023   \pm     0.002   }$\\
&       2014.07.11      &       56849.876       &       -1.88   &       730     &       750     &       60.09   &        -99.51  &       12      &       0.387   $_\mathrm{\pm   0.033   \pm     0.002   }$\\
&       2014.07.11      &       56849.931       &       -0.57   &       660     &       680     &       63.80   &        -115.21 &       12      &       0.262   $_\mathrm{\pm   0.022   \pm     0.002   }$\\
&       2014.07.11      &       56849.944       &       -0.24   &       660     &       680     &       64.40   &        -119.09 &       8       &       0.286   $_\mathrm{\pm   0.036   \pm     0.002   }$\\
&       2014.07.11      &       56849.972       &       0.45    &       610     &       630     &       65.30   &        -127.45 &       7       &       0.269   $_\mathrm{\pm   0.037   \pm     0.003   }$\\
&       2014.07.11      &       56849.984       &       0.72    &       610     &       630     &       65.52   &        -130.73 &       6       &       0.177   $_\mathrm{\pm   0.027   \pm     0.002   }$\\
&       2014.07.11      &       56849.984       &       0.72    &       630     &       650     &       65.52   &        -130.73 &       6       &       0.225   $_\mathrm{\pm   0.039   \pm     0.002   }$\\
\hline
$\phi=0.401\pm0.003$
&       2012.08.29      &       56168.980       &       3.91    &       700     &       720     &       65.51   &                -172.09 &       22      &       0.290   $_\mathrm{\pm    0.013  \pm     0.002   }$\\
&       2012.08.29      &       56168.980       &       3.91    &       720     &       740     &       65.51   &                -172.09 &       29      &       0.362   $_\mathrm{\pm    0.012  \pm     0.003   }$\\
&       2012.08.29      &       56169.013       &        4.70           &       700     &       720     &       65.47   &                177.14  &       18      &       0.309           $_\mathrm{\pm    0.017   \pm     0.002   }$\\
&       2012.08.29      &       56169.013       &        4.70           &       720     &       740     &       65.47   &                177.14  &       24      &       0.379           $_\mathrm{\pm    0.016   \pm     0.003   }$\\
\hline
$\phi=0.526\pm0.001$
&       2013.07.23   &       56497.003     &  1.90    &   525  &    540  &    65.87  &   -145.43 &     3  &    0.094   $_\mathrm{\pm     0.036   \pm 0.001 }$\\
&       2013.07.23   &       56497.003     &  1.90     &    540   &   555  &    65.87  &   -145.43  &    12   &   0.133   $_\mathrm{\pm    0.011 \pm  0.002 }$\\
&       2013.07.23   &       56497.003      &1.90       & 710     & 730   &   65.87   &  -145.43    &  11    &   0.307 $_\mathrm{\pm    0.029  \pm   0.002 }$\\
&       2014.07.07      &       56845.819       &       -3.51   &       514     &       527     &       52.67   &        -78.48  &       9       &       0.307   $_\mathrm{\pm   0.036   \pm     0.003   }$\\
&       2014.07.07      &       56845.819       &       -3.51   &       527     &       542     &       52.67   &        -78.48  &       7       &       0.264   $_\mathrm{\pm   0.038   \pm     0.002   }$\\
&       2014.07.07      &       56845.819       &       -3.51   &       695     &       715     &       52.66   &        -78.46  &       14      &       0.508   $_\mathrm{\pm   0.036   \pm     0.003   }$\\
&       2014.07.07      &       56845.831       &       -3.26   &       514     &       527     &       53.95   &        -81.85  &       8       &       0.241   $_\mathrm{\pm   0.029   \pm     0.002   }$\\
&       2014.07.07      &       56845.831       &       -3.26   &       527     &       542     &       53.95   &        -81.85  &       10      &       0.268   $_\mathrm{\pm   0.026   \pm     0.002   }$\\
&       2014.07.07      &       56845.831       &       -3.26   &       695     &       715     &       53.94   &        -81.84  &       15      &       0.454   $_\mathrm{\pm   0.029   \pm     0.002   }$\\
\hline
$\phi=0.627\pm0.017$
&       2013.10.23      &       56588.841       &       4.12    &       710     &       730     &       65.50   &        -174.91 &       10      &       0.377   $_\mathrm{\pm   0.039   \pm     0.003   }$\\
&       2013.10.23      &       56588.841       &       4.12    &       725     &       740     &       65.50   &        -174.91 &       10      &       0.361   $_\mathrm{\pm   0.038   \pm     0.003   }$\\
&       2013.10.23      &       56588.877       &       4.91    &       710     &       730     &       65.50   &        174.27  &       9       &       0.331   $_\mathrm{\pm   0.036   \pm     0.003   }$\\
&       2013.10.23      &       56588.877       &       4.91    &       725     &       740     &       65.50   &        174.27  &       14      &       0.367   $_\mathrm{\pm   0.027   \pm     0.003   }$\\
&       2014.07.02      &       56840.912       &       -1.51   &       580     &       595     &       61.35   &        -103.99 &       4       &       0.197   $_\mathrm{\pm   0.050   \pm     0.002   }$\\
&       2014.07.02      &       56840.912       &       -1.51   &       730     &       750     &       61.35   &        -103.98 &       7       &       0.354   $_\mathrm{\pm   0.050   \pm     0.002   }$\\
&       2014.07.02      &       56840.930       &       -1.13   &       580     &       595     &       62.46   &        -108.48 &       4       &       0.207   $_\mathrm{\pm   0.050   \pm     0.002   }$\\
&       2014.07.02      &       56840.930       &       -1.13   &       730     &       750     &       62.46   &        -108.47 &       7       &       0.349   $_\mathrm{\pm   0.050   \pm     0.002   }$\\

\end{tabular}
\end{center}
\end{table*}

\begin{table*}
\begin{center}
\caption[]{Observing log with the E1E2 baseline (part 2).}
\label{Tab.log2}
\setlength{\doublerulesep}{\arrayrulewidth}
\begin{tabular}{lccccccccc}
\hline \noalign{\smallskip}
\hline\hline\hline\hline
  &    Date             &  RJD    & HA  &  $\lambda_\mathrm{min}$ & $\lambda_\mathrm{max}$ & Bp &  PA    &   S/N &$V^{2}_\mathrm{cal \pm stat \pm syst}$ \\
                       & [ yyyy.mm.dd ]&  [ days ] & [ hour ] &             [ nm ]&                     [ nm ]  &    [ m ] & [ deg ] &          &                                                         \\
\hline                                                                                                                                                                           
$\phi=0.741\pm0.003$
&       2012.09.27      &       56197.642       &       -2.40   &       532     &       547     &       58.02   &        -93.06  &       11      &       0.272   $_\mathrm{\pm   0.024   \pm     0.003   }$\\
&       2012.09.27      &       56197.642       &       -2.40   &       547     &       560     &       58.02   &        -93.06  &       12      &       0.277   $_\mathrm{\pm   0.022   \pm     0.003   }$\\
&       2012.09.27      &       56197.642       &       -2.42   &       710     &       730     &       57.92   &        -92.79  &       33      &       0.459   $_\mathrm{\pm   0.014   \pm     0.003   }$\\ 
&       2012.09.27      &       56197.642       &       -2.42   &       730     &       750     &       57.92   &        -92.79  &       28      &       0.516   $_\mathrm{\pm   0.018   \pm     0.003   }$\\ 
&       2012.09.27      &       56197.665       &       -1.79   &       532     &       547     &       60.39   &        -100.54 &       14      &       0.280   $_\mathrm{\pm   0.020   \pm     0.003   }$\\
&       2012.09.27      &       56197.665       &       -1.79   &       547     &       560     &       60.39   &        -100.54 &       17      &       0.297   $_\mathrm{\pm   0.018   \pm     0.003   }$\\
&       2012.09.27      &       56197.665       &       -1.80   &       710     &       730     &       60.37   &        -100.47 &       30      &       0.459   $_\mathrm{\pm   0.015   \pm     0.003   }$\\ 
&       2012.09.27      &       56197.665       &       -1.80   &       730     &       750     &       60.37   &        -100.47 &       19      &       0.507   $_\mathrm{\pm   0.026   \pm     0.003   }$\\ 
\hline                                                                                                                                                                           
$\phi=0.783\pm0.002$
&       2013.10.24      &       56589.603       &       -1.61   &       547     &       562     &       61.03   &        -102.77 &       12      &       0.275   $_\mathrm{\pm   0.022   \pm     0.003   }$\\
&       2013.10.24      &       56589.603       &       -1.61   &       705     &       725     &       61.03   &        -102.77 &       18      &       0.424   $_\mathrm{\pm   0.023   \pm     0.003   }$\\
&       2013.10.24      &       56589.603       &       -1.61   &       710     &       730     &       61.03   &        -102.77 &       26      &       0.430   $_\mathrm{\pm   0.016   \pm     0.003   }$\\
&       2013.10.24      &       56589.603       &       -1.61   &       725     &       740     &       61.03   &        -102.77 &       18      &       0.509   $_\mathrm{\pm   0.028   \pm     0.003   }$\\
&       2013.10.24      &       56589.603       &       -1.61   &       725     &       745     &       61.03   &        -102.77 &       22      &       0.461   $_\mathrm{\pm   0.021   \pm     0.003   }$\\
&       2013.10.24      &       56589.615       &       -1.35   &       532     &       547     &       61.83   &        -105.83 &       5       &       0.228   $_\mathrm{\pm   0.050   \pm     0.003   }$\\
&       2013.10.24      &       56589.615       &       -1.35   &       705     &       725     &       61.83   &        -105.83 &       15      &       0.468   $_\mathrm{\pm   0.031   \pm     0.003   }$\\
&       2013.10.24      &       56589.615       &       -1.35   &       710     &       730     &       61.83   &        -105.83 &       24      &       0.470   $_\mathrm{\pm   0.019   \pm     0.003   }$\\
&       2013.10.24      &       56589.615       &       -1.35   &       725     &       740     &       61.83   &        -105.83 &       15      &       0.471   $_\mathrm{\pm   0.031   \pm     0.003   }$\\
&       2013.10.24      &       56589.615       &       -1.35   &       725     &       745     &       61.83   &        -105.83 &       18      &       0.461   $_\mathrm{\pm   0.026   \pm     0.003   }$\\
&       2013.10.24      &       56589.625       &       -1.07   &       532     &       547     &       62.64   &        -109.25 &       7       &       0.211   $_\mathrm{\pm   0.030   \pm     0.002   }$\\
&       2013.10.24      &       56589.625       &       -1.07   &       705     &       725     &       62.64   &        -109.25 &       16      &       0.473   $_\mathrm{\pm   0.029   \pm     0.003   }$\\
&       2013.10.24      &       56589.625       &       -1.07   &       710     &       730     &       62.64   &        -109.25 &       14      &       0.463   $_\mathrm{\pm   0.033   \pm     0.003   }$\\
&       2013.10.24      &       56589.625       &       -1.07   &       725     &       740     &       62.64   &        -109.25 &       13      &       0.426   $_\mathrm{\pm   0.034   \pm     0.003   }$\\
&       2013.10.24      &       56589.625       &       -1.07   &       725     &       745     &       62.64   &        -109.25 &       17      &       0.456   $_\mathrm{\pm   0.027   \pm     0.003   }$\\
\hline                                                                                                                                                                           
$\phi=0.821 \pm0.012$
&       2014.10.24      &       56954.655       &       -0.29   &       570     &       585     &       64.32   &        -118.48 &       11      &       0.238   $_\mathrm{\pm   0.021   \pm     0.003   }$\\
&       2014.10.24      &       56954.655       &       -0.29   &       730     &       750     &       64.32   &        -118.47 &       13      &       0.336   $_\mathrm{\pm   0.025   \pm     0.002   }$\\
&       2014.10.24      &       56954.678       &       0.24    &       520     &       535     &       65.07   &        -124.85 &       7       &       0.145   $_\mathrm{\pm   0.022   \pm     0.002   }$\\
&       2014.10.24      &       56954.678       &       0.24    &       695     &       710     &       65.07   &        -124.84 &       11      &       0.374   $_\mathrm{\pm   0.033   \pm     0.003   }$\\
&       2014.10.24      &       56954.699       &       0.75    &       660     &       680     &       65.54   &        -131.04 &       17      &       0.297   $_\mathrm{\pm   0.018   \pm     0.002   }$\\
&       2014.10.24      &       56954.724       &       1.35    &       610     &       630     &       65.82   &        -138.49 &       8       &       0.221   $_\mathrm{\pm   0.027   \pm     0.002   }$\\
\hline                                                                                                                                                                           
$\phi=0.848 \pm0.007$
&       2013.12.17      &       56643.578       &       1.34    &       532     &       547     &       65.81   &        -138.43 &       6       &       0.224   $_\mathrm{\pm   0.040   \pm     0.003   }$\\
&       2013.12.17      &       56643.578       &       1.34    &       535     &       555     &       65.81   &        -138.43 &       11      &       0.210   $_\mathrm{\pm   0.020   \pm     0.003   }$\\
&       2013.12.17      &       56643.578       &       1.34    &       705     &       725     &       65.81   &        -138.41 &       16      &       0.321   $_\mathrm{\pm   0.020   \pm     0.002   }$\\
&       2013.12.17      &       56643.578       &       1.34    &       725     &       745     &       65.81   &        -138.41 &       14      &       0.388   $_\mathrm{\pm   0.027   \pm     0.003   }$\\
&       2013.12.17      &       56643.592       &       1.65    &       532     &       547     &       65.87   &        -142.23 &       6       &       0.239   $_\mathrm{\pm   0.041   \pm     0.003   }$\\
&       2013.12.17      &       56643.592       &       1.65    &       535     &       555     &       65.87   &        -142.23 &       6       &       0.194   $_\mathrm{\pm   0.031   \pm     0.003   }$\\
&       2013.12.17      &       56643.592       &       1.64    &       705     &       725     &       65.87   &        -142.21 &       11      &       0.274   $_\mathrm{\pm   0.026   \pm     0.002   }$\\
&       2013.12.17      &       56643.592       &       1.64    &       725     &       745     &       65.87   &        -142.21 &       14      &       0.359   $_\mathrm{\pm   0.026   \pm     0.003   }$\\
&       2013.12.17      &       56643.603       &       1.90    &       705     &       725     &       65.87   &        -145.54 &       9       &       0.304   $_\mathrm{\pm   0.034   \pm     0.002   }$\\
&       2013.12.17      &       56643.614       &       2.16    &       532     &       547     &       65.86   &        -148.81 &       6       &       0.220   $_\mathrm{\pm   0.037   \pm     0.003   }$\\
&       2013.12.17      &       56643.614       &       2.16    &       535     &       555     &       65.86   &        -148.81 &       7       &       0.208   $_\mathrm{\pm   0.030   \pm     0.003   }$\\
&       2013.12.17      &       56643.614       &       2.16    &       705     &       725     &       65.86   &        -148.77 &       8       &       0.332   $_\mathrm{\pm   0.040   \pm     0.003   }$\\
&       2013.12.17      &       56643.614       &       2.16    &       725     &       745     &       65.86   &        -148.77 &       8       &       0.300   $_\mathrm{\pm   0.036   \pm     0.002   }$\\
&       2013.12.17      &       56643.624       &       2.42    &       705     &       725     &       65.82   &        -152.21 &       6       &       0.385   $_\mathrm{\pm   0.062   \pm     0.003   }$\\
&       2013.12.17      &       56643.624       &       2.42    &       725     &       745     &       65.82   &        -152.21 &       6       &       0.379   $_\mathrm{\pm   0.060   \pm     0.003   }$\\
&       2013.12.17      &       56643.644       &       2.91    &       532     &       547     &       65.72   &        -158.69 &       5       &       0.190   $_\mathrm{\pm   0.040   \pm     0.002   }$\\
&       2013.12.17      &       56643.644       &       2.91    &       705     &       725     &       65.72   &        -158.66 &       9       &       0.386   $_\mathrm{\pm   0.042   \pm     0.003   }$\\
&       2013.12.17      &       56643.676       &       3.67    &       535     &       555     &       65.56   &        -168.84 &       8       &       0.203   $_\mathrm{\pm   0.026   \pm     0.003   }$\\
&       2013.12.17      &       56643.676       &       3.67    &       705     &       725     &       65.56   &        -168.81 &       12      &       0.357   $_\mathrm{\pm   0.030   \pm     0.003   }$\\
&       2013.12.17      &       56643.676       &       3.67    &       725     &       745     &       65.56   &        -168.81 &       13      &       0.391   $_\mathrm{\pm   0.031   \pm     0.003   }$\\
&       2013.12.17      &       56643.687       &       3.92    &       532     &       547     &       65.52   &        -172.21 &       8       &       0.266   $_\mathrm{\pm   0.034   \pm     0.003   }$\\
&       2013.12.17      &       56643.687       &       3.92    &       535     &       555     &       65.52   &        -172.21 &       6       &       0.207   $_\mathrm{\pm   0.032   \pm     0.003   }$\\
&       2013.12.17      &       56643.687       &       3.91    &       705     &       725     &       65.52   &        -172.18 &       13      &       0.390   $_\mathrm{\pm   0.029   \pm     0.003   }$\\
&       2013.12.17      &       56643.687       &       3.91    &       725     &       745     &       65.52   &        -172.18 &       15      &       0.370   $_\mathrm{\pm   0.024   \pm     0.003   }$\\

\end{tabular}
\end{center}
\end{table*}

\begin{table*}
\begin{center}
\caption[]{Observing with the E1E2 baseline (part 3).}
\label{Tab.log3}
\setlength{\doublerulesep}{\arrayrulewidth}
\begin{tabular}{lccccccccc}
\hline \noalign{\smallskip}
\hline\hline\hline\hline
   &    Date             &   RJD    & HA  &  $\lambda_\mathrm{min}$ & $\lambda_\mathrm{max}$ & Bp &  PA    &   S/N &$V^{2}_\mathrm{cal \pm stat \pm syst}$ \\
                       & [ yyyy.mm.dd ]&  [ days ] & [ hour ] &             [ nm ]&                     [ nm ]  &    [ m ] & [ deg ] &          &                                                         \\
      \hline
$\phi=0.893 \pm0.005$
&       2013.07.25      &       56498.942       &       0.64    &       525     &       540     &       65.46   &        -129.77 &       8       &       0.174   $_\mathrm{\pm   0.022   \pm     0.003   }$\\
&       2013.07.25      &       56498.942       &       0.64    &       540     &       555     &       65.46   &        -129.77 &       9       &       0.191   $_\mathrm{\pm   0.021   \pm     0.003   }$\\
&       2013.07.25      &       56498.942       &       0.64    &       690     &       710     &       65.46   &        -129.77 &       16      &       0.287   $_\mathrm{\pm   0.017   \pm     0.002   }$\\
&       2013.07.25      &       56498.942       &       0.64    &       710     &       730     &       65.46   &        -129.77 &       15      &       0.302   $_\mathrm{\pm   0.020   \pm     0.002   }$\\
&       2013.07.25      &       56498.942       &       0.64    &       730     &       750     &       65.46   &        -129.77 &       15      &       0.361   $_\mathrm{\pm   0.024   \pm     0.002   }$\\
&       2013.07.25      &       56498.961       &       1.04    &       525     &       540     &       65.71   &        -134.71 &       5       &       0.194   $_\mathrm{\pm   0.037   \pm     0.003   }$\\
&       2013.07.25      &       56498.961       &       1.04    &       540     &       555     &       65.71   &        -134.71 &       7       &       0.247   $_\mathrm{\pm   0.036   \pm     0.003   }$\\
&       2013.07.25      &       56498.961       &       1.04    &       690     &       710     &       65.71   &        -134.70 &       11      &       0.310   $_\mathrm{\pm   0.027   \pm     0.002   }$\\
&       2013.07.25      &       56498.961       &       1.04    &       710     &       730     &       65.71   &        -134.70 &       11      &       0.307   $_\mathrm{\pm   0.027   \pm     0.002   }$\\
&       2013.07.25      &       56498.961       &       1.04    &       730     &       750     &       65.71   &        -134.70 &       11      &       0.273   $_\mathrm{\pm   0.026   \pm     0.002   }$\\
&       2013.07.25      &       56498.978       &       1.41    &       525     &       540     &       65.83   &        -139.32 &       7       &       0.221   $_\mathrm{\pm   0.031   \pm     0.003   }$\\
&       2013.07.25      &       56498.978       &       1.41    &       540     &       555     &       65.83   &        -139.32 &       11      &       0.236   $_\mathrm{\pm   0.022   \pm     0.003   }$\\
&       2013.07.25      &       56498.978       &       1.41    &       690     &       710     &       65.83   &        -139.31 &       17      &       0.299   $_\mathrm{\pm   0.017   \pm     0.002   }$\\
&       2013.07.25      &       56498.978       &       1.41    &       710     &       730     &       65.83   &        -139.31 &       13      &       0.340   $_\mathrm{\pm   0.025   \pm     0.002   }$\\
&       2013.07.25      &       56498.978       &       1.41    &       730     &       750     &       65.83   &        -139.31 &       20      &       0.297   $_\mathrm{\pm   0.015   \pm     0.002   }$\\
&       2013.07.25      &       56498.988       &       1.67    &       525     &       540     &       65.87   &        -142.58 &       9       &       0.176   $_\mathrm{\pm   0.020   \pm     0.002   }$\\
&       2013.07.25      &       56498.988       &       1.67    &       540     &       555     &       65.87   &        -142.58 &       12      &       0.208   $_\mathrm{\pm   0.017   \pm     0.003   }$\\
&       2013.07.25      &       56498.988       &       1.67    &       690     &       710     &       65.87   &        -142.57 &       22      &       0.299   $_\mathrm{\pm   0.014   \pm     0.002   }$\\
&       2013.07.25      &       56498.988       &       1.67    &       710     &       730     &       65.87   &        -142.57 &       15      &       0.337   $_\mathrm{\pm   0.023   \pm     0.002   }$\\
&       2013.07.25      &       56498.988       &       1.67    &       730     &       750     &       65.87   &        -142.57 &       19      &       0.325   $_\mathrm{\pm   0.017   \pm     0.002   }$\\
&       2013.07.25      &       56498.999       &       1.95    &       525     &       540     &       65.87   &        -146.15 &       11      &       0.211   $_\mathrm{\pm   0.020   \pm     0.003   }$\\
&       2013.07.25      &       56498.999       &       1.95    &       540     &       555     &       65.87   &        -146.15 &       12      &       0.238   $_\mathrm{\pm   0.019   \pm     0.003   }$\\
&       2013.07.25      &       56498.999       &       1.95    &       690     &       710     &       65.87   &        -146.14 &       21      &       0.330   $_\mathrm{\pm   0.015   \pm     0.003   }$\\
&       2013.07.25      &       56498.999       &       1.95    &       710     &       730     &       65.87   &        -146.14 &       25      &       0.369   $_\mathrm{\pm   0.015   \pm     0.003   }$\\
&       2013.07.25      &       56498.999       &       1.95    &       730     &       750     &       65.87   &        -146.14 &       18      &       0.337   $_\mathrm{\pm   0.018   \pm     0.002   }$\\
&       2013.07.25      &       56499.010       &       2.16    &       525     &       540     &       65.86   &        -148.78 &       9       &       0.186   $_\mathrm{\pm   0.020   \pm     0.003   }$\\
&       2013.07.25      &       56499.010       &       2.16    &       540     &       555     &       65.86   &        -148.78 &       10      &       0.197   $_\mathrm{\pm   0.020   \pm     0.003   }$\\
&       2013.07.25      &       56499.010       &       2.16    &       690     &       710     &       65.86   &        -148.77 &       12      &       0.288   $_\mathrm{\pm   0.025   \pm     0.002   }$\\
&       2013.07.25      &       56499.010       &       2.16    &       710     &       730     &       65.86   &        -148.77 &       13      &       0.330   $_\mathrm{\pm   0.025   \pm     0.002   }$\\
&       2013.07.25      &       56499.010       &       2.16    &       730     &       750     &       65.86   &        -148.77 &       14      &       0.314   $_\mathrm{\pm   0.022   \pm     0.002   }$\\
\hline
$\phi=0.927 \pm0.003$
&       2014.07.09      &       56847.948       &       -0.29   &       580     &       595     &       64.33   &        -118.57 &       5       &       0.239   $_\mathrm{\pm   0.050   \pm     0.002   }$\\
&       2014.07.09      &       56847.948       &       -0.29   &       735     &       755     &       64.33   &        -118.56 &       7       &       0.368   $_\mathrm{\pm   0.050   \pm     0.002   }$\\
&       2014.07.09      &       56847.961       &       0.03    &       580     &       595     &       64.81   &        -122.35 &       4       &       0.212   $_\mathrm{\pm   0.050   \pm     0.002   }$\\
&       2014.07.09      &       56847.961       &       0.03    &       735     &       755     &       64.81   &        -122.33 &       8       &       0.390   $_\mathrm{\pm   0.050   \pm     0.002   }$\\
&       2014.07.09      &       56847.978       &       0.47    &       490     &       510     &       65.32   &        -127.66 &       3       &       0.173   $_\mathrm{\pm   0.050   \pm     0.003   }$\\
&       2014.07.09      &       56847.978       &       0.47    &       660     &       680     &       65.32   &        -127.65 &       7       &       0.341   $_\mathrm{\pm   0.050   \pm     0.003   }$\\
&       2014.07.09      &       56847.990       &       0.74    &       490     &       510     &       65.54   &        -130.97 &       4       &       0.181   $_\mathrm{\pm   0.050   \pm     0.003   }$\\
&       2014.07.09      &       56847.990       &       0.74    &       660     &       680     &       65.53   &        -130.96 &       6       &       0.307   $_\mathrm{\pm   0.050   \pm     0.003   }$\\
\hline
$\phi=0.942 \pm0.007$
&       2013.11.26      &       56622.628       &       1.23    &       547     &       562     &       65.78   &        -137.04 &       4       &       0.181   $_\mathrm{\pm   0.050   \pm     0.002   }$\\
&       2013.11.26      &       56622.628       &       1.23    &       705     &       725     &       65.78   &        -137.03 &       7       &       0.327   $_\mathrm{\pm   0.050   \pm     0.002   }$\\
&       2013.11.26      &       56622.628       &       1.23    &       725     &       740     &       65.78   &        -137.03 &       6       &       0.316   $_\mathrm{\pm   0.050   \pm     0.002   }$\\
&       2013.11.26      &       56622.644       &       1.58    &       547     &       562     &       65.86   &        -141.35 &       5       &       0.260   $_\mathrm{\pm   0.050   \pm     0.003   }$\\
&       2013.11.26      &       56622.644       &       1.57    &       705     &       725     &       65.86   &        -141.26 &       6       &       0.313   $_\mathrm{\pm   0.050   \pm     0.002   }$\\
&       2013.11.26      &       56622.644       &       1.57    &       725     &       740     &       65.86   &        -141.26 &       7       &       0.363   $_\mathrm{\pm   0.050   \pm     0.003   }$\\
&       2013.11.26      &       56622.658       &       1.90    &       535     &       555     &       65.87   &        -145.46 &       2       &       0.118   $_\mathrm{\pm   0.050   \pm     0.002   }$\\
&       2013.11.26      &       56622.658       &       1.90    &       547     &       562     &       65.87   &        -145.46 &       3       &       0.165   $_\mathrm{\pm   0.050   \pm     0.002   }$\\
&       2013.11.26      &       56622.658       &       1.90    &       705     &       725     &       65.87   &        -145.45 &       6       &       0.284   $_\mathrm{\pm   0.050   \pm     0.003   }$\\
&       2013.11.26      &       56622.658       &       1.90    &       725     &       740     &       65.87   &        -145.45 &       6       &       0.297   $_\mathrm{\pm   0.050   \pm     0.002   }$\\
&       2013.11.26      &       56622.698       &       2.85    &       597     &       617     &       65.73   &        -157.82 &       4       &       0.214   $_\mathrm{\pm   0.050   \pm     0.002   }$\\
&       2013.11.26      &       56622.698       &       2.85    &       617     &       637     &       65.73   &        -157.82 &       5       &       0.248   $_\mathrm{\pm   0.050   \pm     0.002   }$\\
&       2013.11.26      &       56622.712       &       3.17    &       617     &       637     &       65.66   &        -162.08 &       6       &       0.313   $_\mathrm{\pm   0.050   \pm     0.003   }$\\
\hline
$\phi=0.999 \pm0.025$
&       2013.10.25      &       56590.620       &       -1.09   &       705     &       725     &       62.59   &        -109.03 &       16      &       0.430   $_\mathrm{\pm   0.027   \pm     0.003   }$\\
&       2013.10.25      &       56590.620       &       -1.09   &       725     &       745     &       62.59   &        -109.03 &       17      &       0.469   $_\mathrm{\pm   0.027   \pm     0.003   }$\\
&       2013.10.25      &       56590.841       &       4.20    &       532     &       547     &       65.49   &        -176.10 &       10      &       0.189   $_\mathrm{\pm   0.019   \pm     0.002   }$\\
&       2013.10.25      &       56590.841       &       4.20    &       705     &       725     &       65.49   &        -176.09 &       19      &       0.372   $_\mathrm{\pm   0.019   \pm     0.003   }$\\
&       2013.10.25      &       56590.841       &       4.20    &       725     &       745     &       65.49   &        -176.09 &       20      &       0.426   $_\mathrm{\pm   0.022   \pm     0.003   }$\\
&       2013.10.25      &       56590.860       &       4.68    &       532     &       547     &       65.49   &        177.37  &       3       &       0.195   $_\mathrm{\pm   0.024   \pm     0.003   }$\\
&       2013.10.25      &       56590.860       &       4.68    &       705     &       725     &       65.49   &        177.37  &       14      &       0.346   $_\mathrm{\pm   0.024   \pm     0.003   }$\\
&       2013.10.25      &       56590.860       &       4.68    &       725     &       745     &       65.49   &        177.37  &       18      &       0.382   $_\mathrm{\pm   0.025   \pm     0.003   }$\\
\end{tabular}
\end{center}
\end{table*}

\begin{table*}
\begin{center}
\caption[]{Observing log with the S1S2 baseline.}
\label{Tab.log4}
\setlength{\doublerulesep}{\arrayrulewidth}
\begin{tabular}{lccccccccc}
\hline \noalign{\smallskip}
\hline\hline\hline\hline
   &    Date             &   RJD    & HA  &  $\lambda_\mathrm{min}$ & $\lambda_\mathrm{max}$ & Bp &  Arg    &   S/N &$V^{2}_\mathrm{cal \pm stat \pm syst}$ \\
                       & [ yyyy.mm.dd ]&  [ days ] & [ h ] &             [ nm ]&                     [ nm ]  &    [ m ] & [ deg ] &          &                                                         \\
 \hline
 $\phi=0.300 $
&       2008.08.03      &       54681.988       &       2.36    &       628     &       632     &       29.01   &        148.13  &       13      &       0.649   $_\mathrm{\pm   0.037   \pm     0.003   }$\\
&       2008.08.03      &       54681.988       &       2.36    &       654     &       658     &       29.01   &        148.13  &       15      &       0.763   $_\mathrm{\pm   0.049   \pm     0.003   }$\\
 \hline
$\phi=0.543 \pm 0.009$
&       2008.07.30      &       54677.894       &       -0.19   &       629     &       632     &       30.97   &        171.02  &       8       &       0.667   $_\mathrm{\pm   0.079   \pm     0.003   }$\\
&       2008.07.30      &       54677.894       &       -0.19   &       654     &       658     &       30.97   &        171.01  &       11      &       0.694   $_\mathrm{\pm   0.063   \pm     0.003   }$\\
&       2008.07.30      &       54677.958       &       1.37    &       530     &       533     &       30.08   &        156.84  &       12      &       0.615   $_\mathrm{\pm   0.049   \pm     0.004   }$\\
&       2008.07.30      &       54677.958       &       1.37    &       553     &       558     &       30.08   &        156.84  &       13      &       0.629   $_\mathrm{\pm   0.048   \pm     0.004   }$\\
 \hline
$\phi=0.666 \pm 0.006 $
&       2008.08.05      &       54683.928       &       1.10    &       554     &       559     &       30.29   &        159.25  &       12      &       0.619   $_\mathrm{\pm   0.030   \pm     0.004   }$\\
&       2008.08.05      &       54683.974       &       2.18    &       507     &       510     &       29.24   &        149.68  &       6       &       0.544   $_\mathrm{\pm   0.096   \pm     0.004   }$\\
\hline
$\phi=0.821 \pm0.012$
&       2014.10.24      &       56954.792       &       2.99    &       570     &       585     &       28.10   &        142.69  &       20      &       0.655   $_\mathrm{\pm   0.033   \pm     0.001   }$\\
&       2014.10.24      &       56954.792       &       2.99    &       730     &       750     &       28.10   &        142.70  &       23      &       0.752   $_\mathrm{\pm   0.033   \pm     0.001   }$\\
&       2014.10.24      &       56954.812       &       3.49    &       695     &       710     &       27.25   &        138.53  &       24      &       0.720   $_\mathrm{\pm   0.030   \pm     0.001   }$\\
\hline
\end{tabular}
\end{center}
\end{table*}

\begin{figure*}[htbp]
\begin{center}
\resizebox{0.9\hsize}{!}{\includegraphics[clip=true]{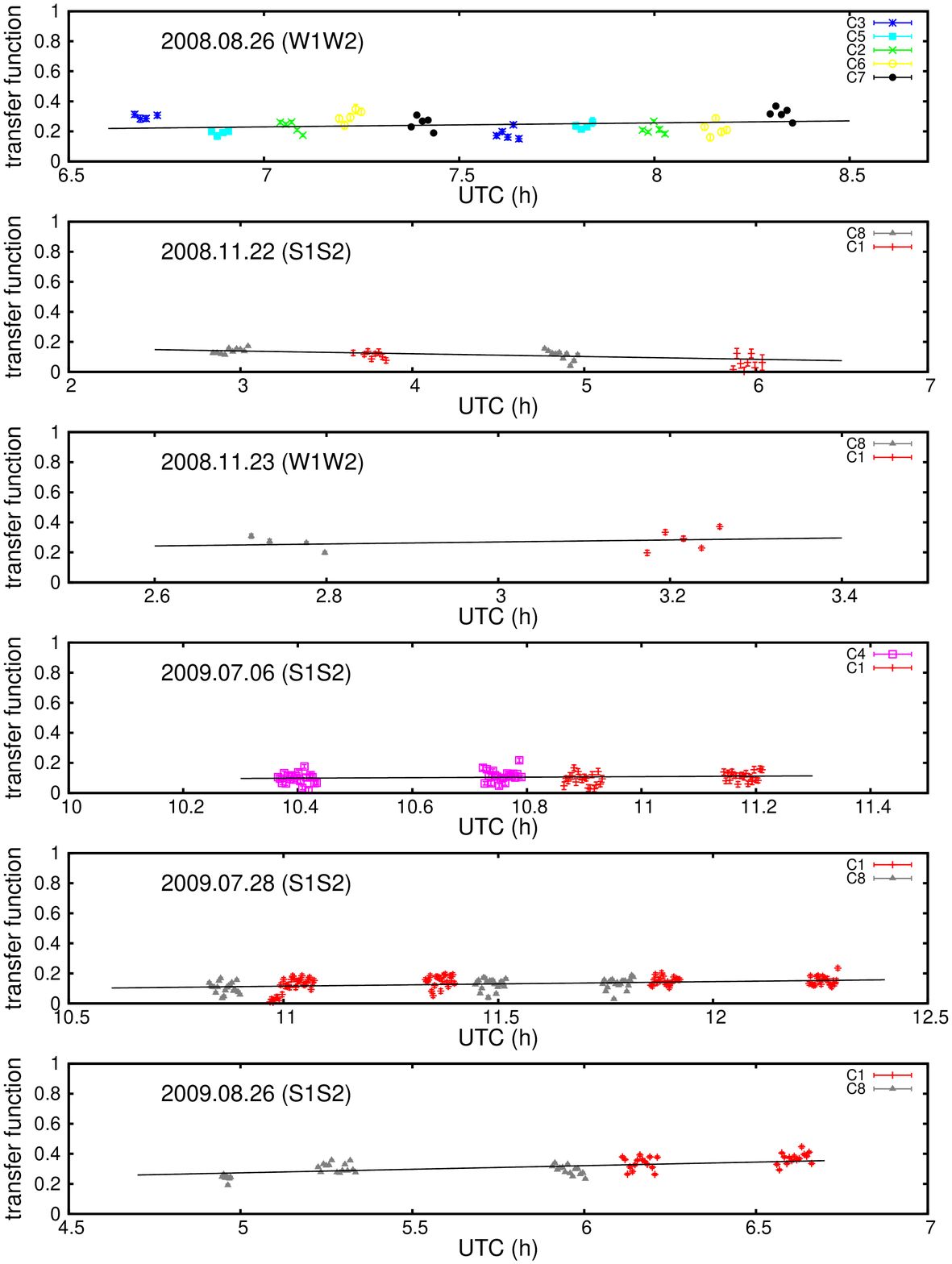}}
\end{center}
\vspace{-3cm}
\caption{Two calibrators, C1 and C2, used to calibrate the VEGA/CHARA data of $\delta$~Cep are compared in terms of transfer functions with six other calibrators, C3 to C8 (listed in Tab. \ref{Tab.cals}), on six different nights during 2008 and 2009. The consistency between the calibrators, which is shown by a linear fit (black line), is independent of the two-telescope baseline considered, S1S2 or W1W2.} \label{Fig.cals}
\end{figure*}

\begin{figure*}[htbp]
\begin{center}
\resizebox{0.9\hsize}{!}{\includegraphics[clip=true]{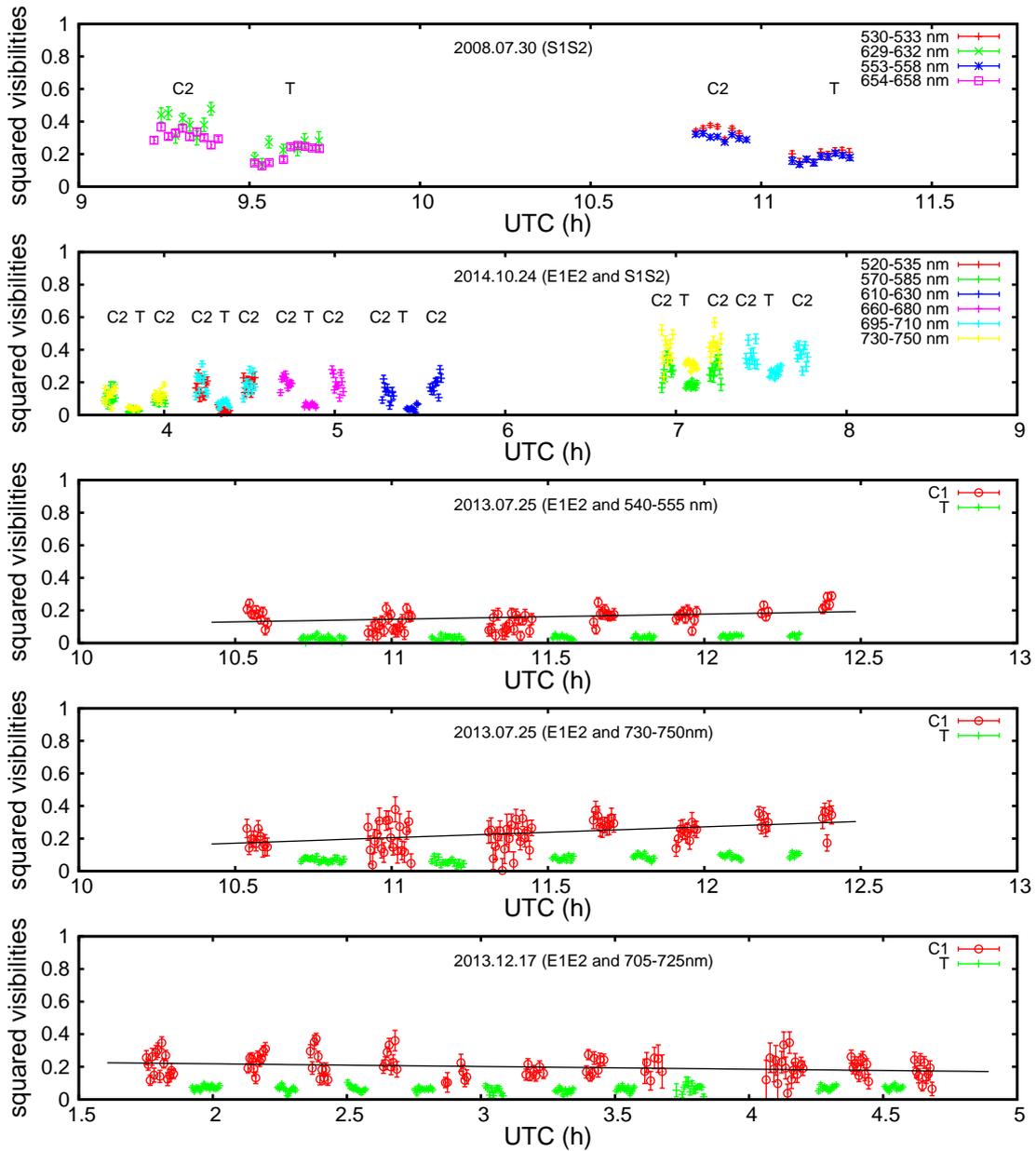}}
\end{center}
\vspace{-6cm}
\caption{Transfer function (calculated either from the C1 or C2 calibrator) is shown together with the raw visibilities of the target  (T) for four different nights in our sample, using different spectral configurations and two baselines, E1E2 and S1S2. For nights 2008 July 30 and 2014 October 24, the squared visibilities (or transfer functions) associated with  calibrators C1 and C2 cannot be consistently compared as they correspond to different spectral configurations, but they are shown to illustrate the quality of the data. The 2014 October 24 observing night is the only one for which we have successive data from E1E2 (from around 3.5h to 6h in UTC) and S1S2 (from around 6.5h to 8h in UTC). For nights 2013 July 25 and 2013 December 17, the instrument configuration is the same over   the entire observing time, and    the stability of the transfer function is evident (black line).} \label{Fig.ft}
\end{figure*}

\begin{figure*}[htbp]
\begin{center}
\resizebox{0.35\hsize}{!}{\includegraphics[clip=true]{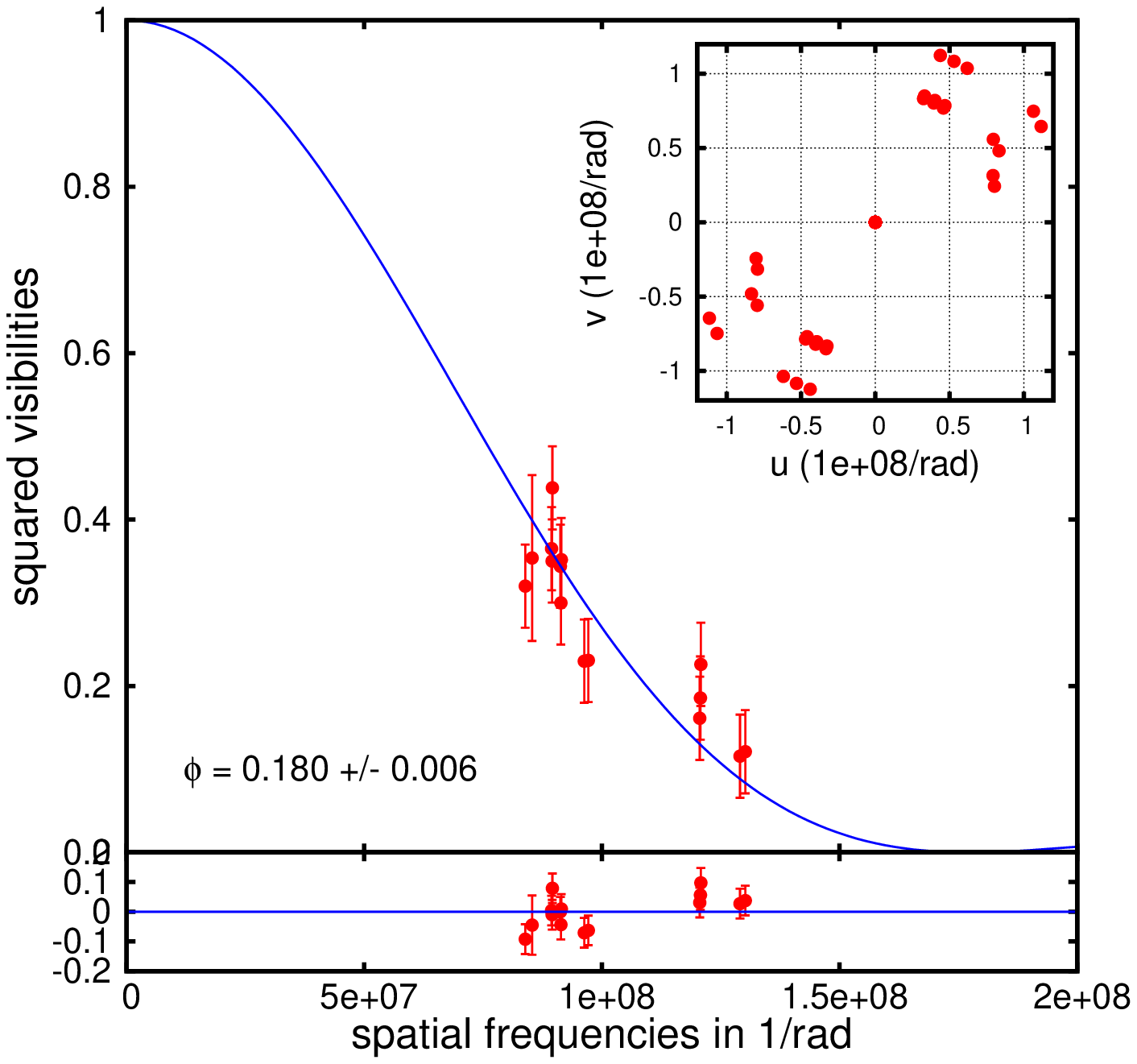}}
\resizebox{0.35\hsize}{!}{\includegraphics[clip=true]{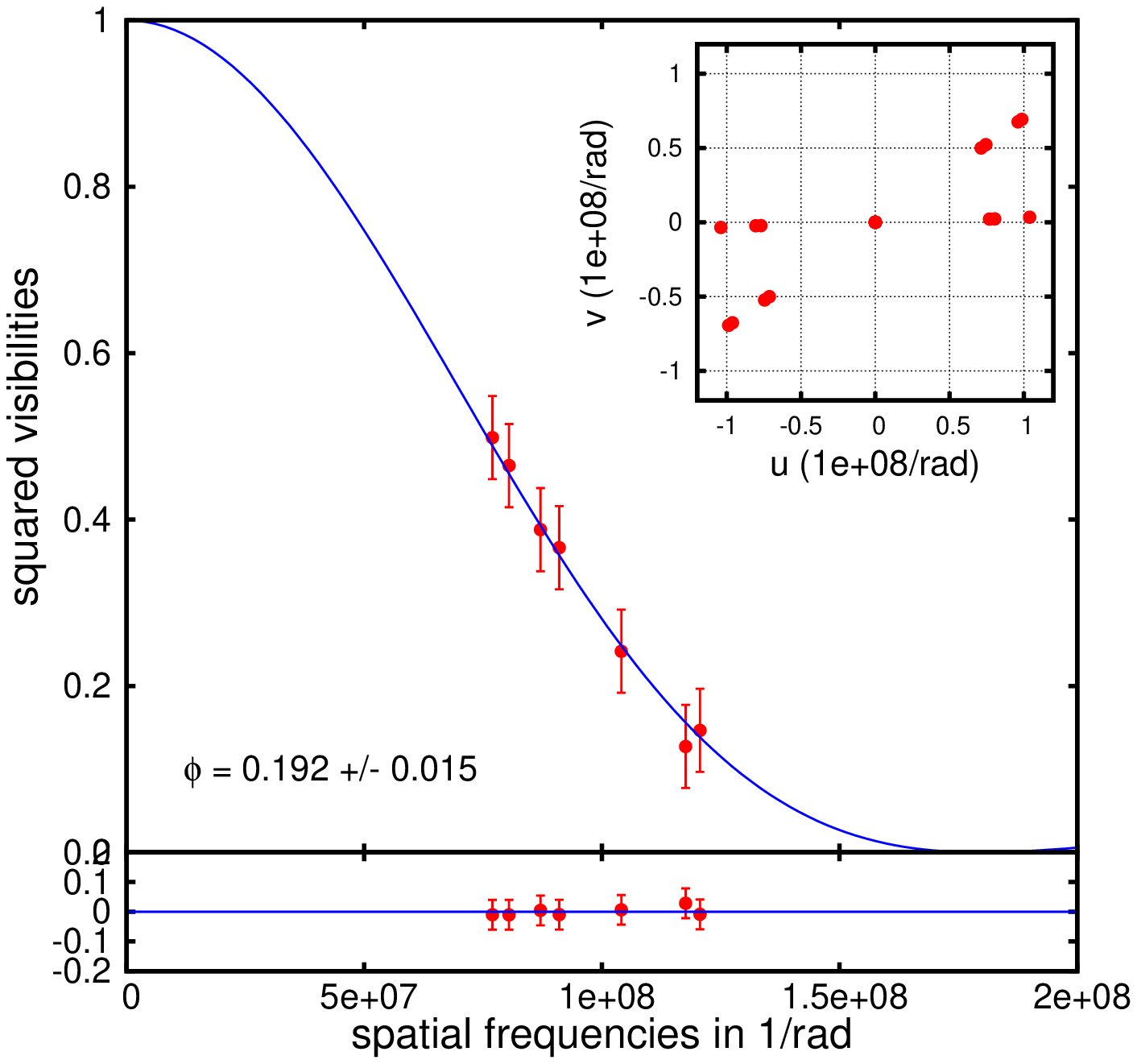}}
\end{center}
\begin{center}
\resizebox{0.30\hsize}{!}{\includegraphics[clip=true]{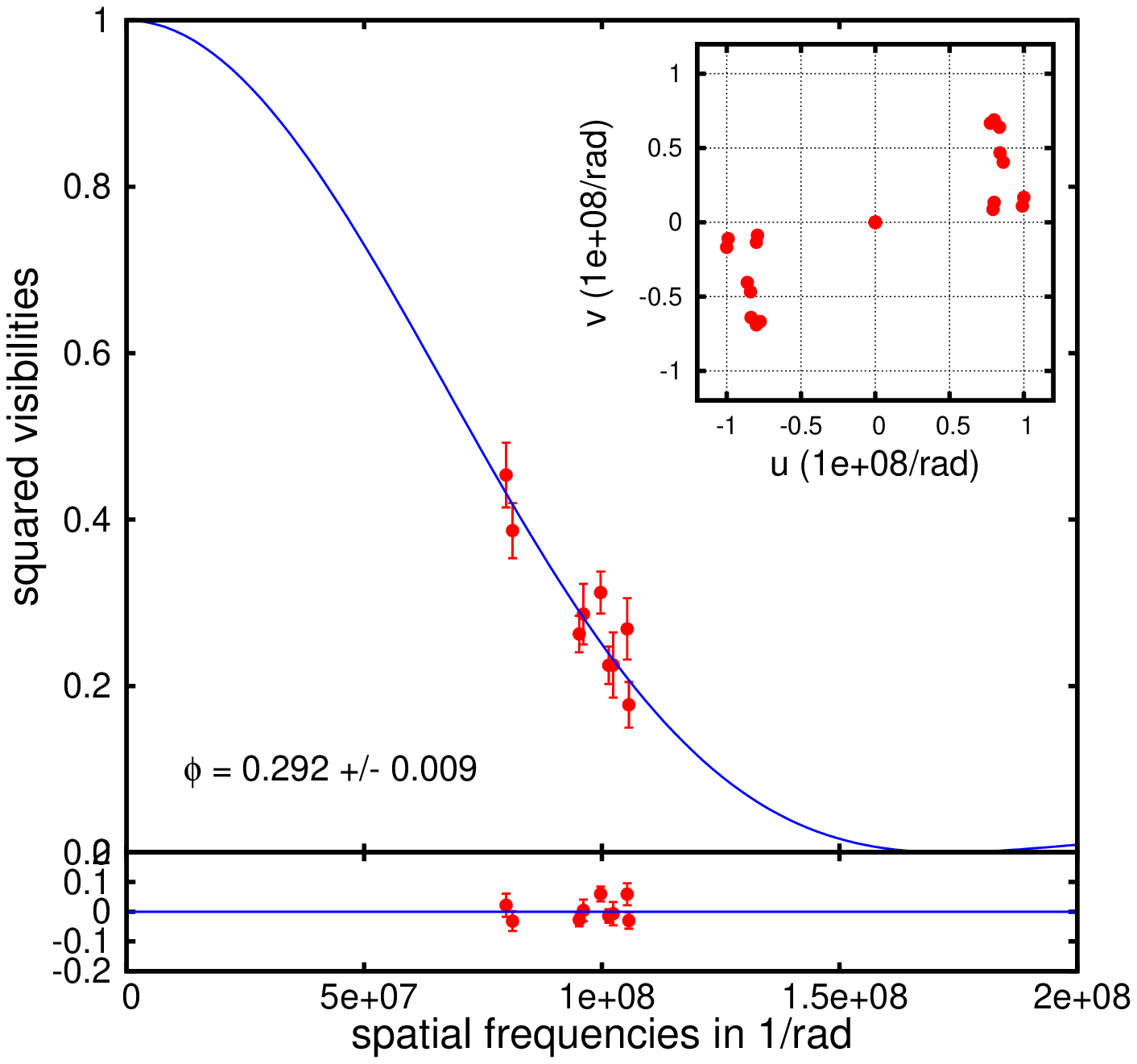}}
\resizebox{0.30\hsize}{!}{\includegraphics[clip=true]{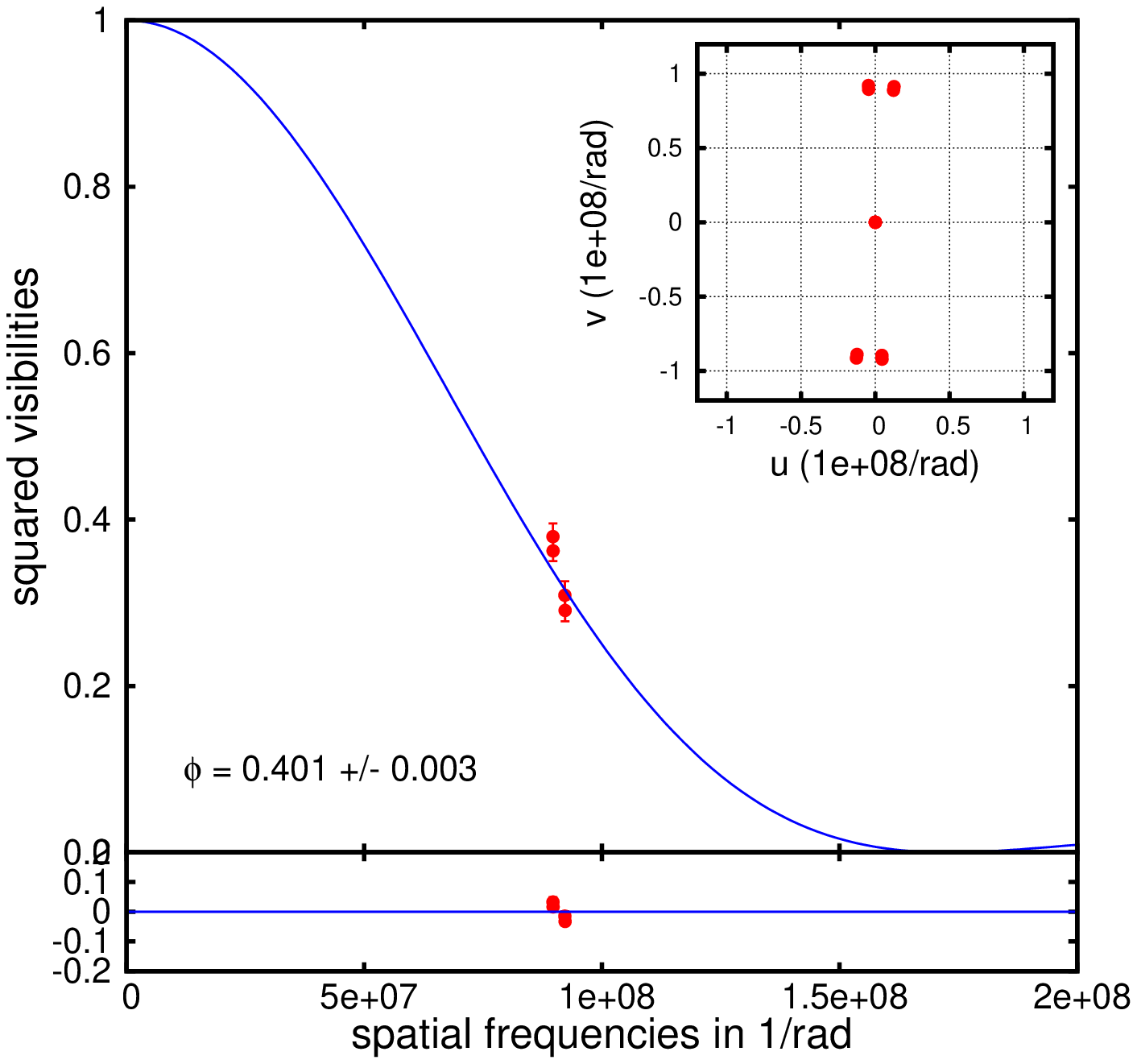}}
\resizebox{0.30\hsize}{!}{\includegraphics[clip=true]{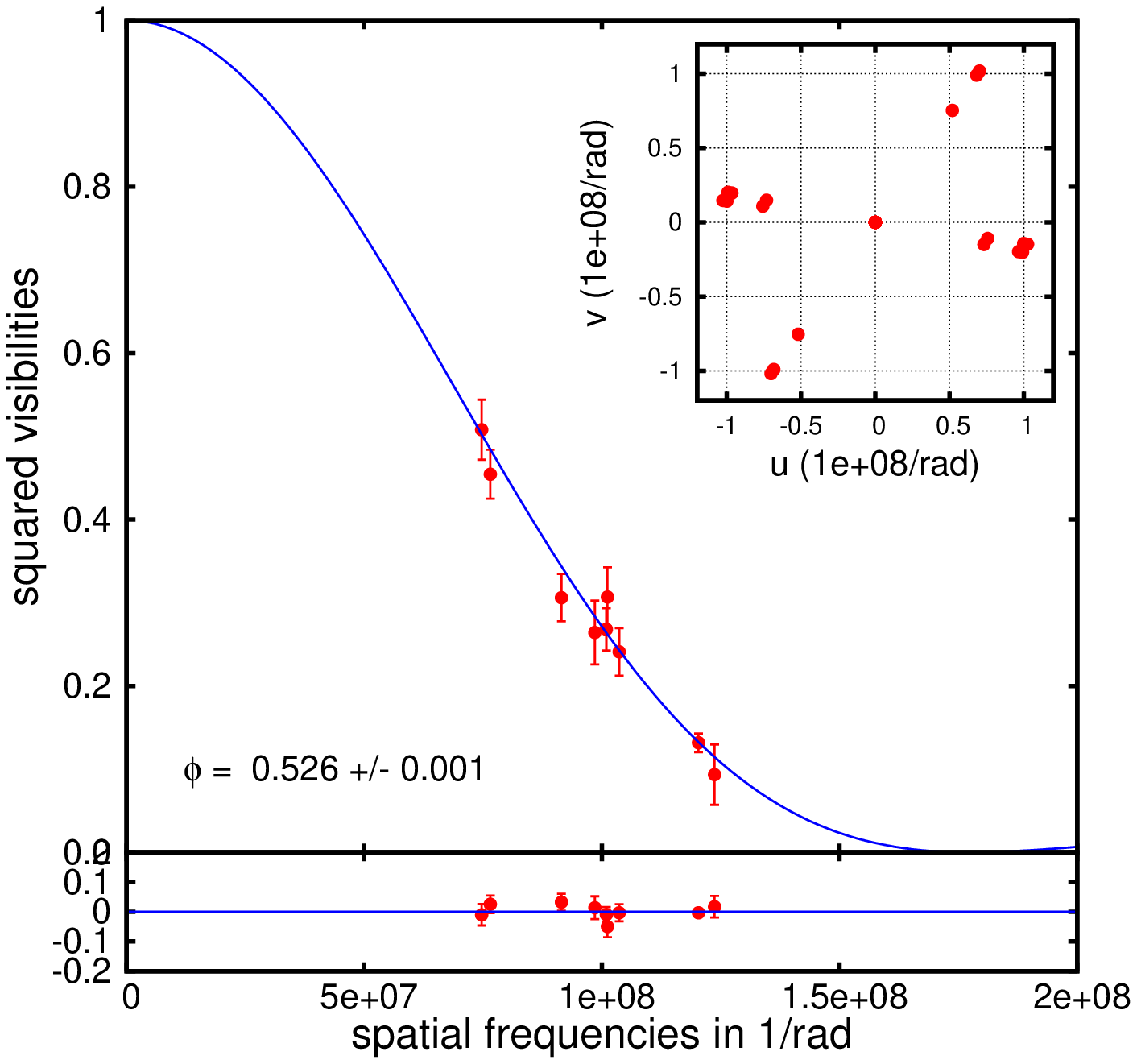}}
\end{center}
\begin{center}
\resizebox{0.30\hsize}{!}{\includegraphics[clip=true]{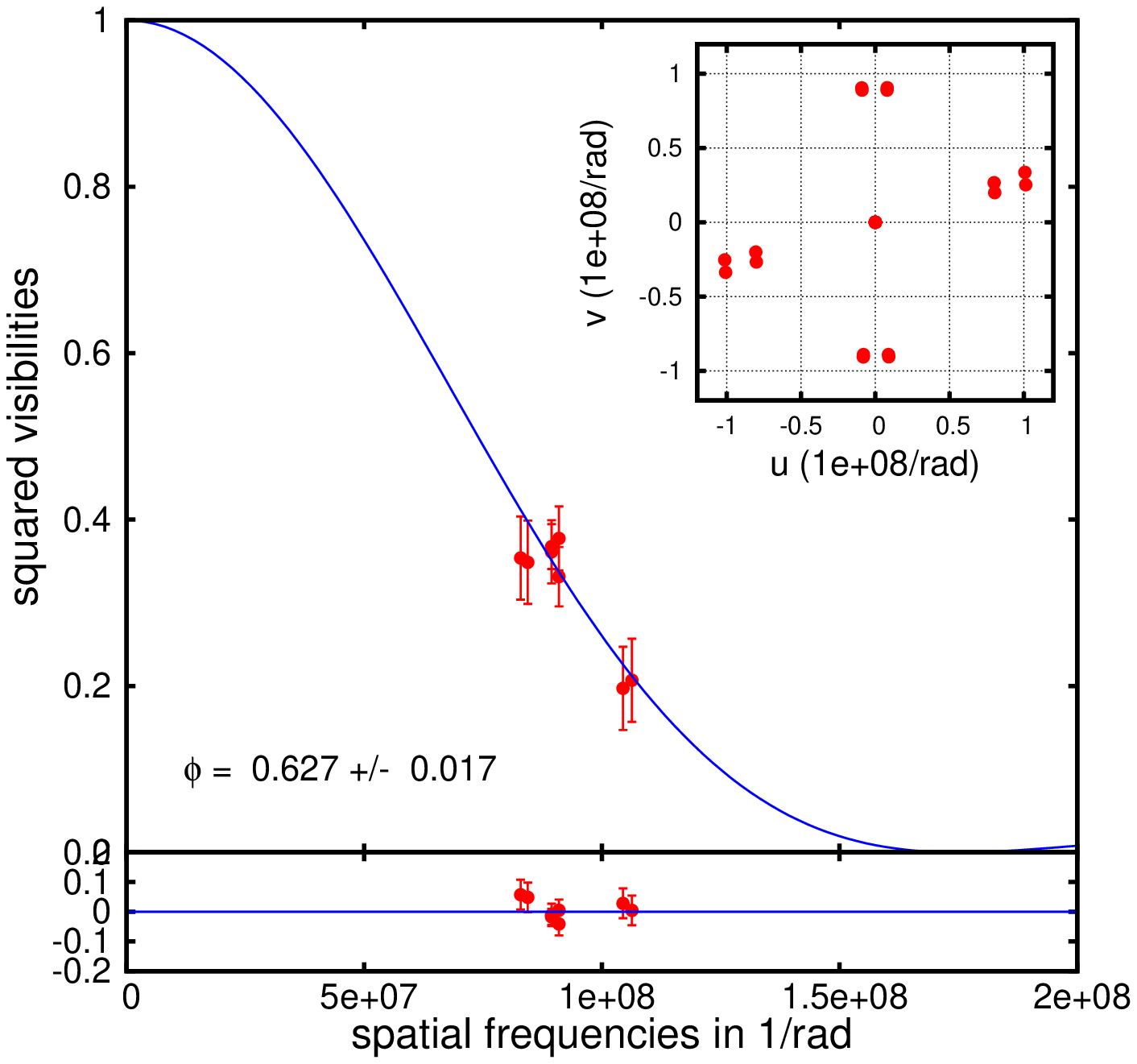}}
\resizebox{0.30\hsize}{!}{\includegraphics[clip=true]{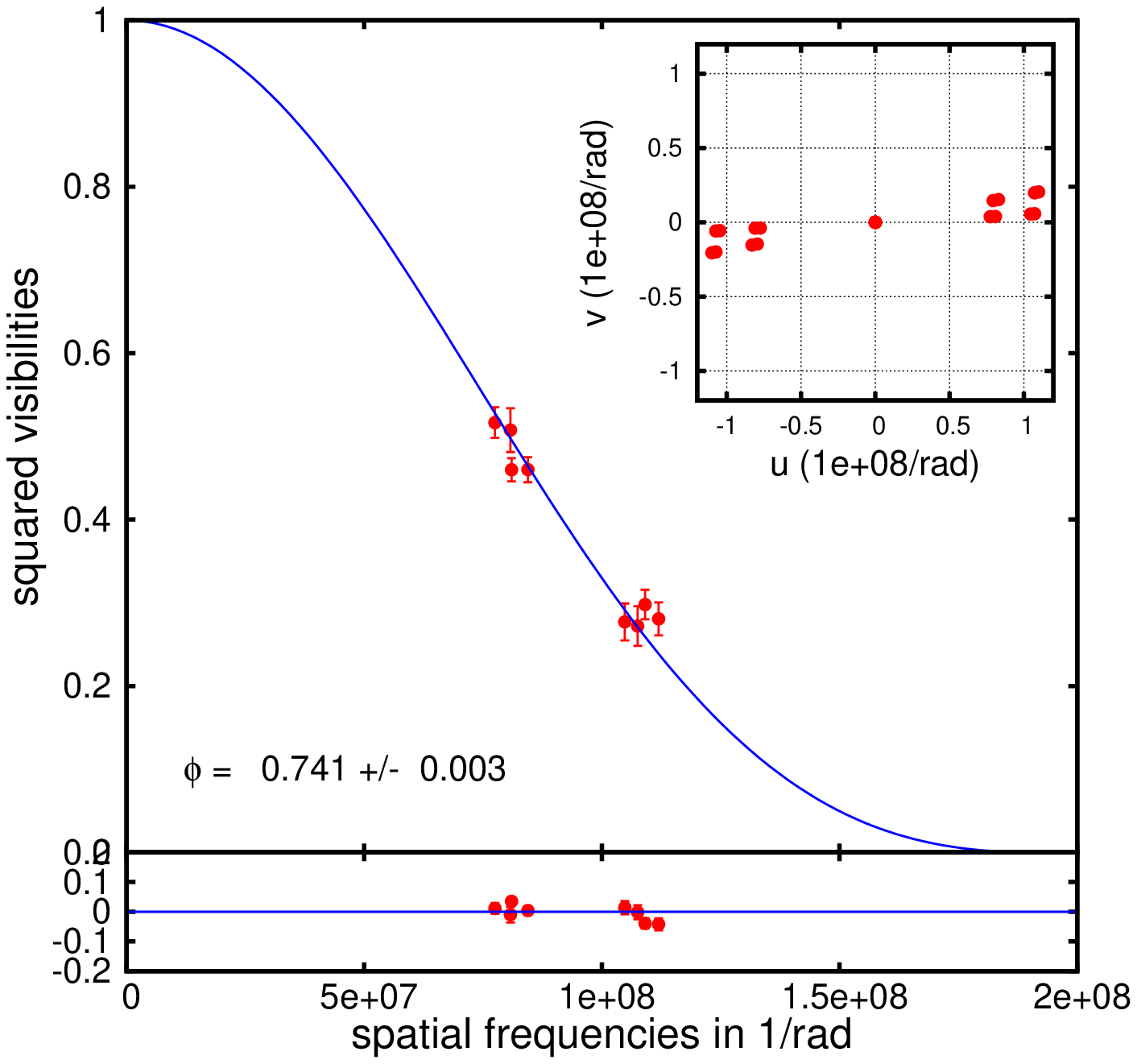}}
\resizebox{0.30\hsize}{!}{\includegraphics[clip=true]{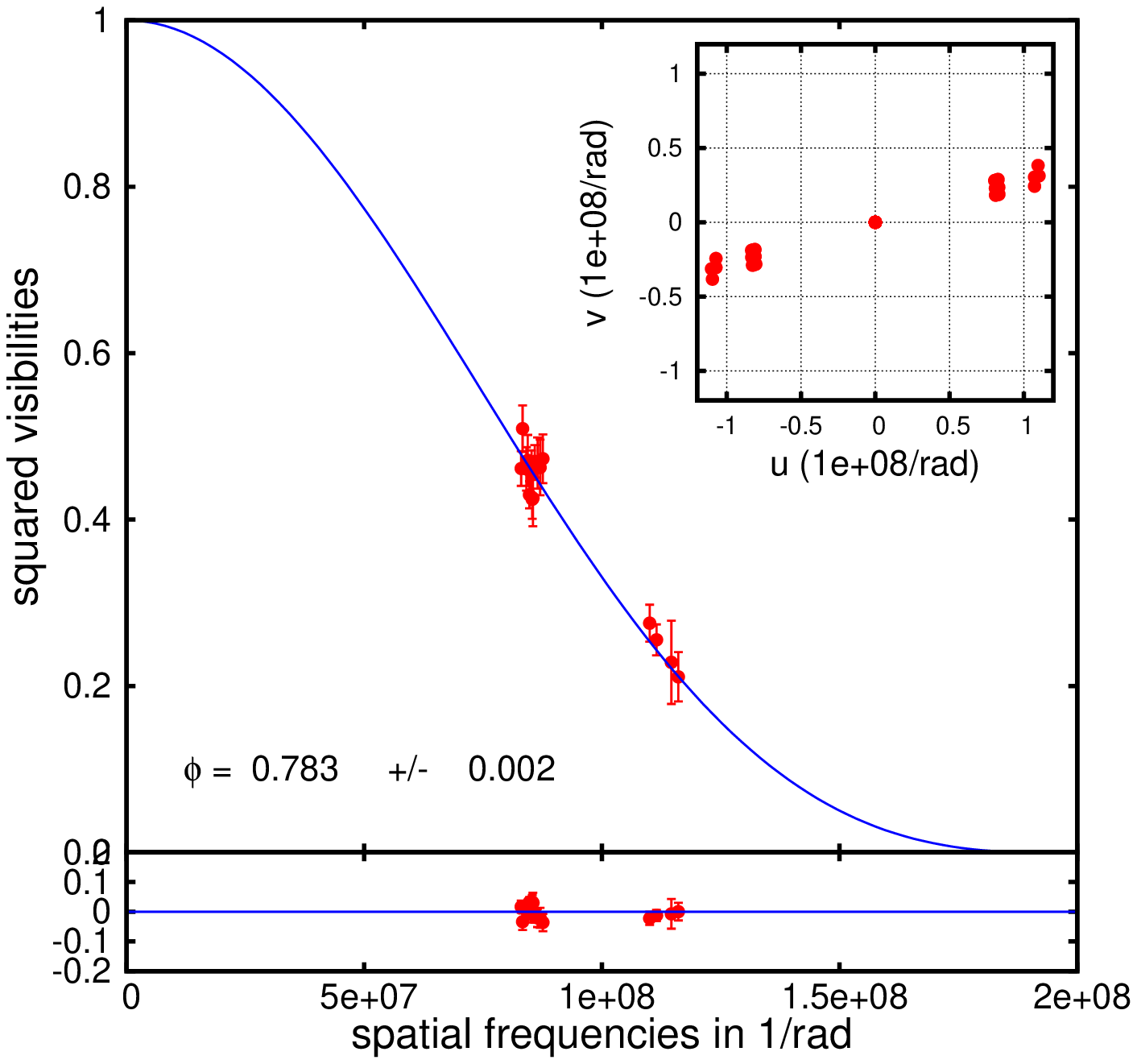}}
\end{center}
\caption{ Observed calibrated squared visibilities (red dots) are plotted as a function of the spatial frequency for each pulsation phase (indicated in the lower left corner of each panel) together with the best fit of uniform disk model (solid blue line). In the upper right corner we show the corresponding (u,v) coverage. } \label{Fig.all1}
\end{figure*}

\begin{figure*}[htbp]
\begin{center}
\resizebox{0.35\hsize}{!}{\includegraphics[clip=true]{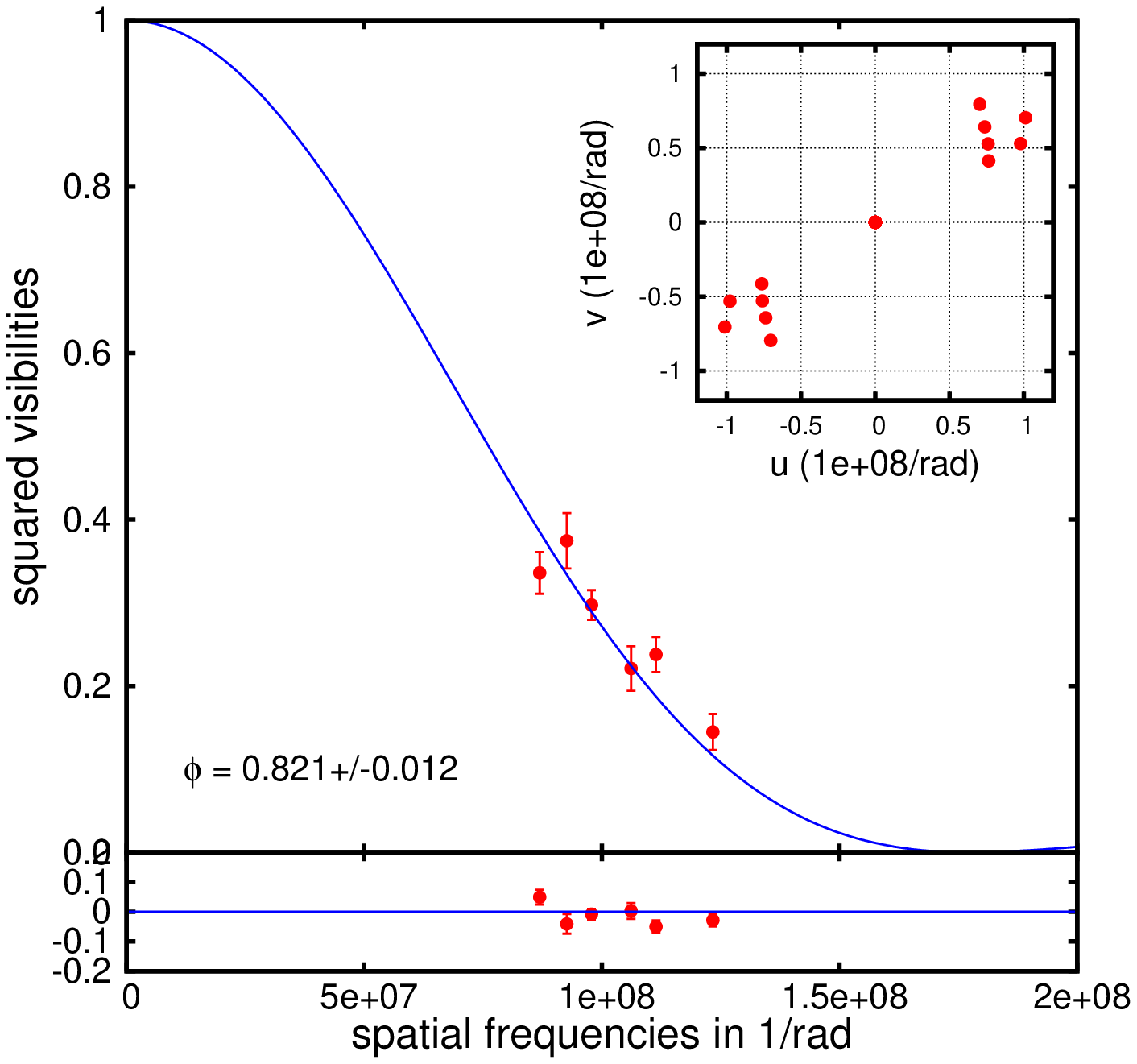}}
\resizebox{0.35\hsize}{!}{\includegraphics[clip=true]{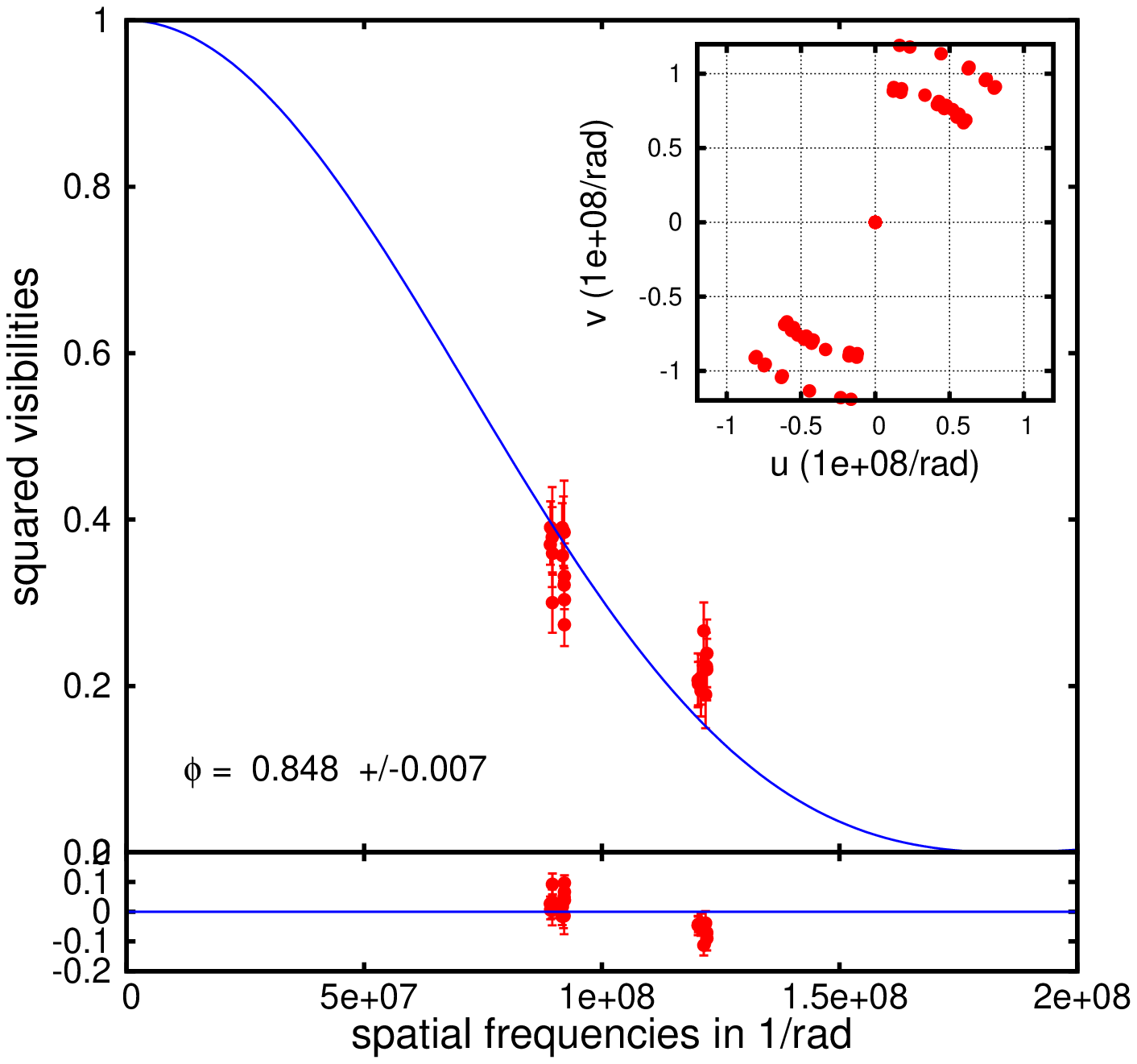}}
\end{center}
\begin{center}
\resizebox{0.35\hsize}{!}{\includegraphics[clip=true]{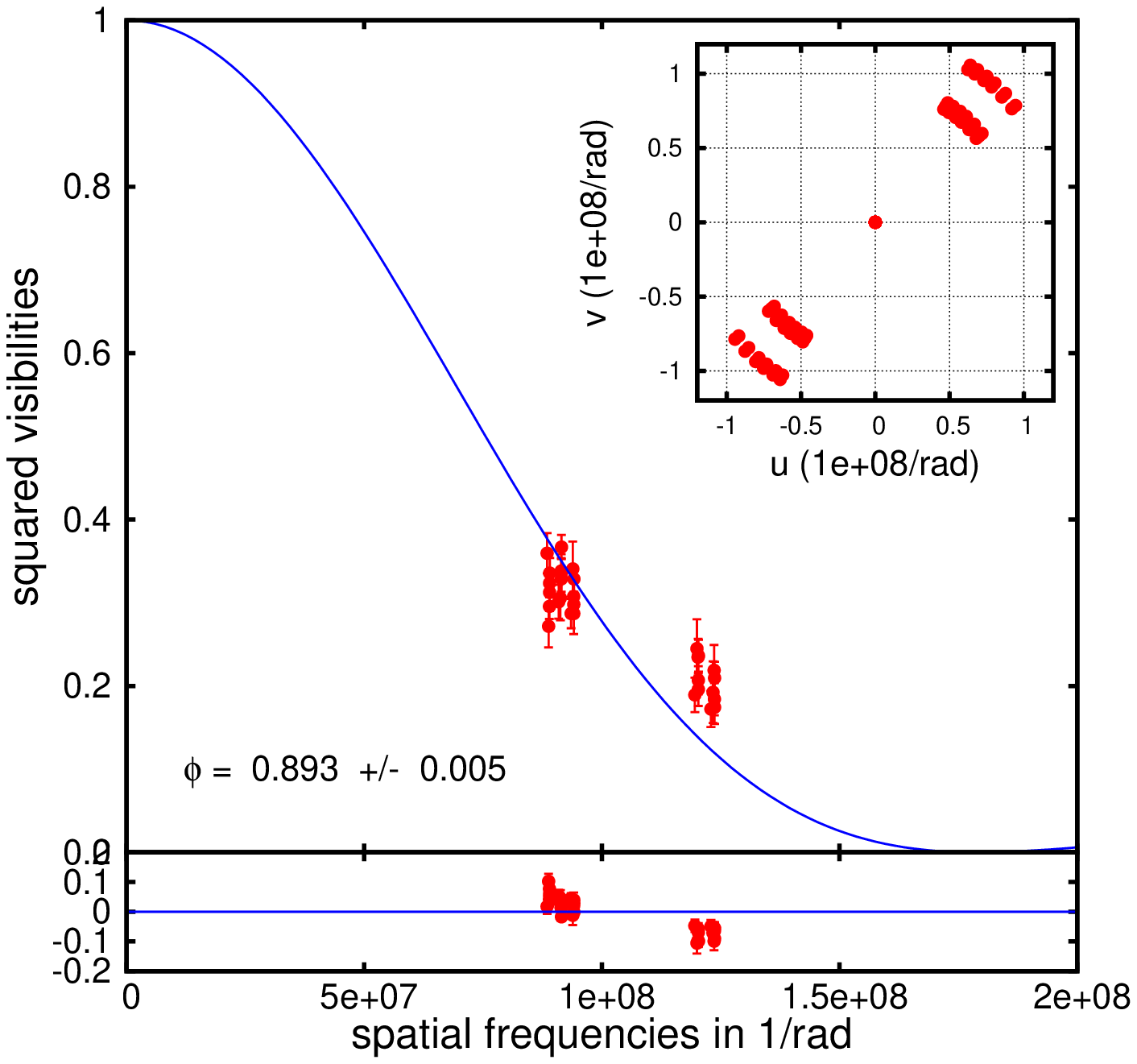}}
\resizebox{0.35\hsize}{!}{\includegraphics[clip=true]{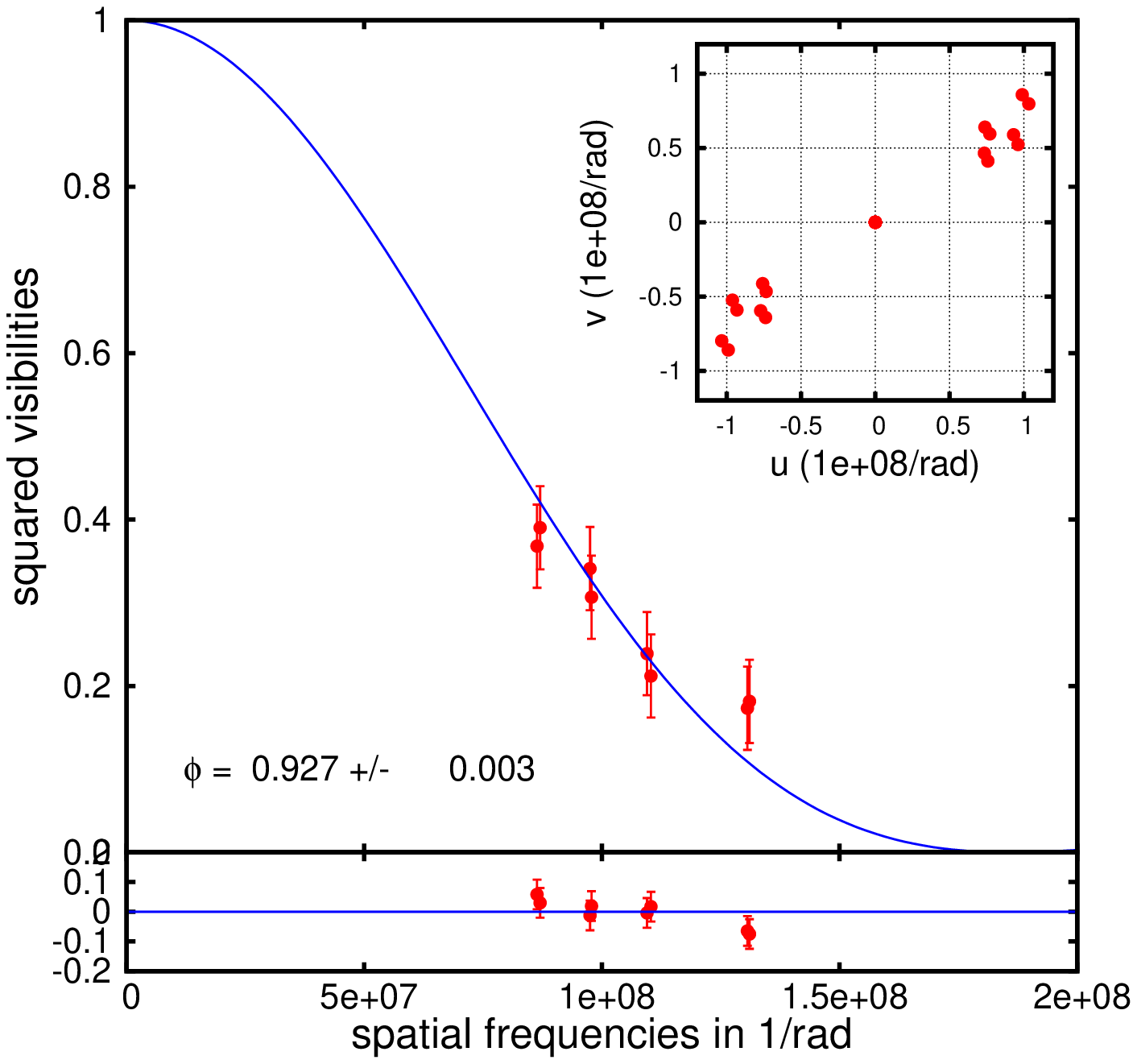}}
\end{center}
\begin{center}
\resizebox{0.35\hsize}{!}{\includegraphics[clip=true]{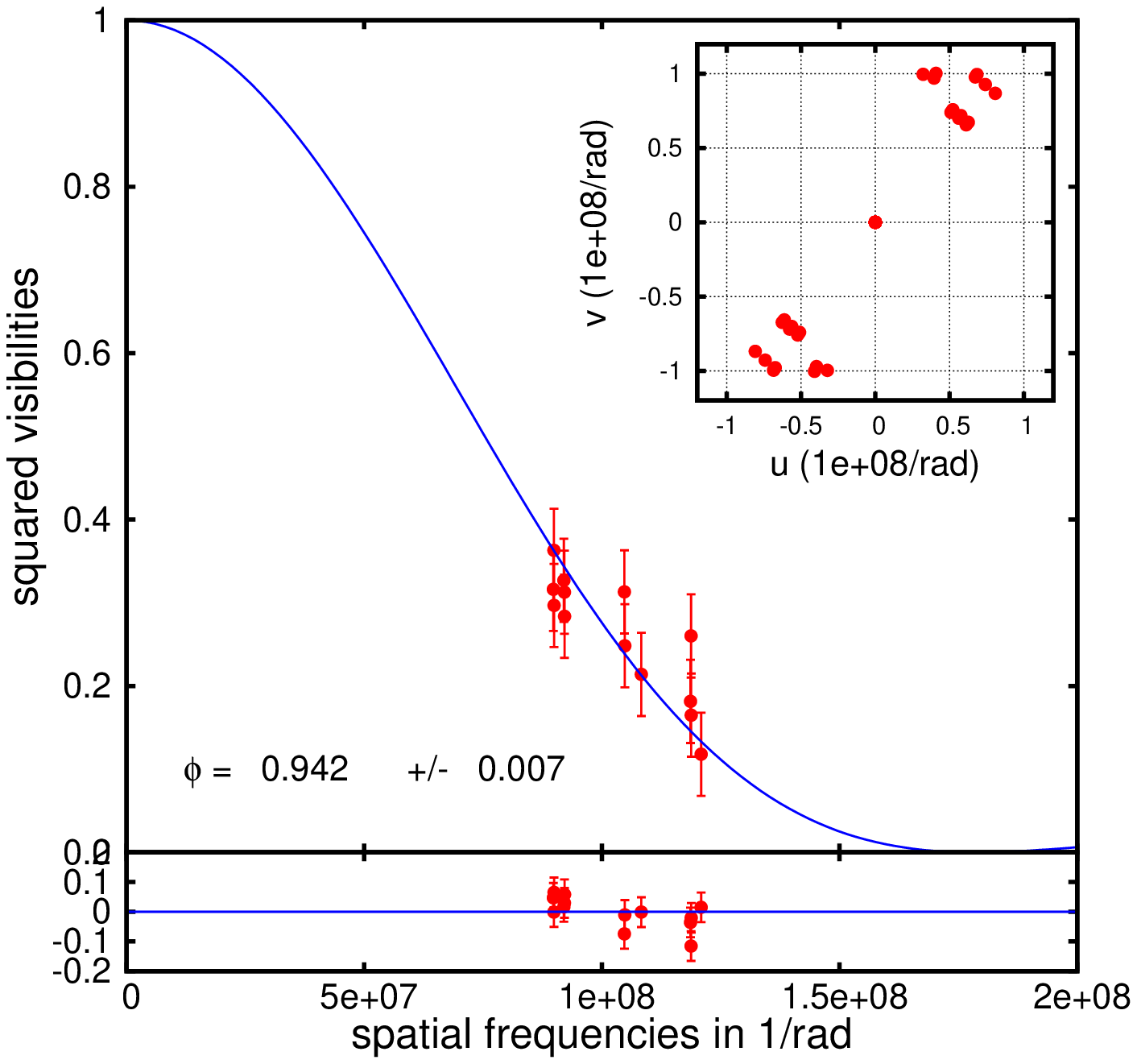}}
\resizebox{0.35\hsize}{!}{\includegraphics[clip=true]{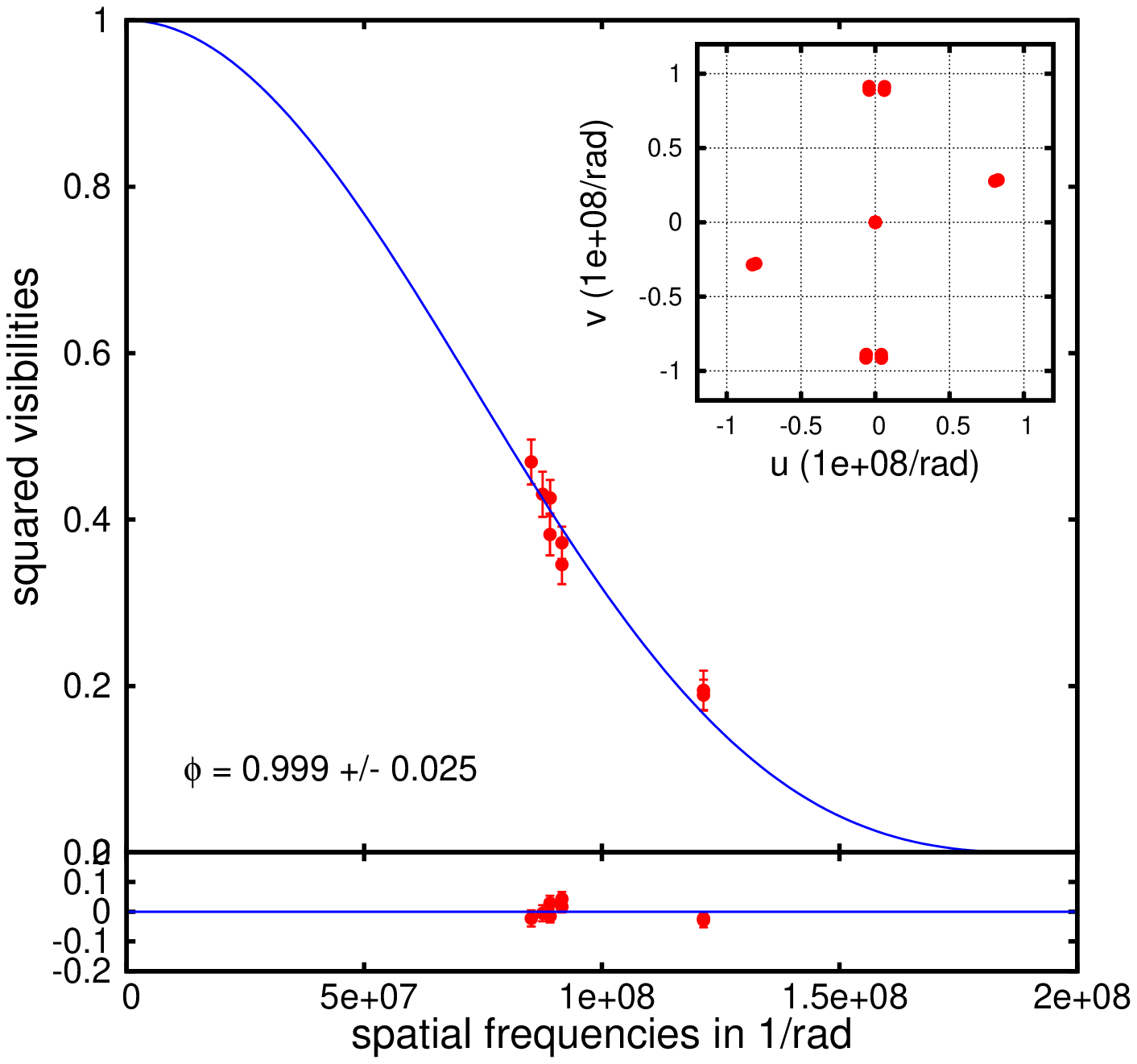}}
\end{center}
\caption{Fig. \ref{Fig.all1} continuation.} \label{Fig.all2}
\end{figure*}

\end{appendix}
\end{document}